\documentclass[pdflatex,sn-mathphys-num]{sn-jnl}

\usepackage{graphicx}%
\usepackage{multirow}%
\usepackage{amsmath,amssymb,amsfonts}%
\usepackage{amsthm}%
\usepackage{mathrsfs}%
\usepackage[title]{appendix}%
\usepackage{xcolor}%
\usepackage{textcomp}%
\usepackage{manyfoot}%
\usepackage{booktabs}%
\usepackage{algorithm}%
\usepackage{algorithmicx}%
\usepackage{algpseudocode}%
\usepackage{listings}%
\usepackage{amsmath}
\usepackage{array}
\usepackage{caption}
\usepackage{enumitem}
\usepackage{subcaption}
\usepackage[utf8]{inputenc}
\usepackage{enumitem} 

\usepackage{booktabs} 
\usepackage{array} 
\usepackage{siunitx} 

\theoremstyle{thmstyleone}%
\theoremstyle{thmstyletwo}%

\theoremstyle{thmstylethree}%

\begin{document}

\title[Article Title]{MeD-3D: A Multimodal Deep Learning Framework for Precise Recurrence Prediction in Clear Cell Renal Cell Carcinoma (ccRCC)}


\author*[1]{\fnm{Hasaan} \sur{Maqsood}}\email{hasaan.maqsood@skoltech.ru}

\author*[2]{\fnm{Saif Ur Rehman} \sur{Khan}}\email{saif\_ur\_rehman.Khan@dfki.de}
\equalcont{These authors contributed equally to this work.}

\affil*[1]{
  \orgdiv{} 
  \orgname{Skolkovo Institute of Science and Technology (Skoltech)}, \\ \orgaddress{\street{Bolshoy Boulevard, 30, bld.1}, \city{Moscow}, 
  \postcode{121205}, 
  \country{Russia}
  }
}

\affil*[2]{
  \orgdiv{} 
  \orgname{German Research Center for Artificial Intelligence (DFKI)}, 
  \orgaddress{
    \street{Trippstadter Str. 122}, 
    \city{Kaiserslautern}, 
    \postcode{67663}, 
    \country{Germany}
  }
}







\abstract{
\textbf{Purpose:} 
Accurate prediction of recurrence in clear cell renal cell carcinoma (ccRCC) remains a major clinical challenge due to the disease complex molecular, pathological, and clinical heterogeneity. Traditional prognostic models, which rely on single data modalities such as radiology, histopathology, or genomics, often fail to capture the full spectrum of disease complexity, resulting in suboptimal predictive accuracy. This study aims to overcome these limitations by proposing a deep learning (DL) framework that integrates multimodal data, including CT, MRI, histopathology whole slide images (WSI), clinical data, and genomic profiles, to improve the prediction of ccRCC recurrence and enhance clinical decision-making.

\textbf{Method:} 
The proposed framework utilizes a comprehensive dataset curated from multiple publicly available sources, including TCGA, TCIA, and CPTAC. To process the diverse modalities, domain-specific models are employed: CLAM, a ResNet50-based model, is used for histopathology WSIs, while MeD-3D, a pre-trained 3D-ResNet18 model, processes CT and MRI images. For structured clinical and genomic data, a multi-layer perceptron (MLP) is used. These models are designed to extract deep feature embeddings from each modality, which are then fused through an early and late integration architecture. This fusion strategy enables the model to combine complementary information from multiple sources. Additionally, the framework is designed to handle incomplete data, a common challenge in clinical settings, by enabling inference even when certain modalities are missing.

\textbf{Results:} 
Experimental validation demonstrates that the proposed MeD-3D model significantly outperforms unimodal baselines across a range of performance metrics. Notably, the MeD-3D model shows superior accuracy in sparse data scenarios, where certain modalities are missing or incomplete, emphasizing the strength of combining multiple data types to capture the full complexity of ccRCC recurrence. The MeD-3D model also provides a more robust and generalized approach compared to traditional methods, improving the predictive performance in clinical settings where data availability is often limited.

\textbf{Conclusion:} 
This study presents a robust and scalable MeD-3D multimodal DL pipeline that integrates diverse biomedical data sources for ccRCC recurrence prediction. By leveraging the complementary information from CT, MRI, histopathology, clinical data, and genomic profiles, the proposed approach significantly enhances prediction accuracy and risk stratification. The framework offers direct implications for personalized treatment planning, enabling clinicians to better tailor interventions based on individual patient profiles. Furthermore, this work enhance the application of DL in precision oncology by addressing common issues such as incomplete data and demonstrating the utility of multimodal integration for improving clinical outcomes in ccRCC.}


%
%
%




\maketitle

\section{Introduction}\label{sec1}
Cancer recurrence remains one of the most significant challenges in modern oncology, posing considerable obstacles to effective long-term treatment and patient survival. Despite advances in early detection, targeted therapies, and immunotherapies, the recurrence of cancer, particularly in metastatic or resistant forms, continues to complicate clinical management. The ability of cancer cells to evade treatment through mechanisms such as genetic mutations, immune evasion, and tumor microenvironment adaptations contributes to the persistence and recurrence of the disease. Moreover, cancer continues to be a leading cause of morbidity and mortality worldwide, with more than 19 million new cases and nearly 10 million deaths recorded in 2020 alone \cite{who2020}. Among the most critical challenges in oncology today is the problem of cancer recurrence where a tumor returns after initial treatment often in a more aggressive or treatment-resistant form. Recurrence significantly reduces survival prospects, increases treatment complexity, and burdens healthcare systems. As oncology enters the era of precision medicine, effectively predicting recurrence remains a major unresolved challenge. Traditional unimodal approaches often fail to capture the full complexity of cancer, which spans diverse clinical, imaging, and molecular data sources. Recent advances in deep neural networks particularly multimodal fusion using Graph Neural Networks and Transformers offer promising avenues for more accurate and personalized prediction by integrating heterogeneous cancer data at multiple scales  \cite{waqas2024multimodal}.

Traditionally, recurrence risk has been assessed using unimodal prognostic models that rely on clinical staging systems, radiological imaging, or molecular biomarkers in isolation. However, cancer is a multifaceted disease, characterized by heterogeneity at the spatial, cellular, and molecular levels. These complexities are not adequately captured by single-modality approaches. Consequently, conventional models often fall short in identifying high-risk patients who may benefit from early adjuvant therapy or enhanced surveillance strategies.

\subsection{Clear Cell Renal Cell Carcinoma: A Case for Multimodal Precision Prognostics}
ccRCC is the most common and biologically aggressive subtype of kidney cancer, accounting for approximately 75--80\% of all renal malignancies \cite{linehan2019cancer}. It is characterized histologically by clear cytoplasm filled with glycogen and lipids, and molecularly by biallelic inactivation of the \textit{VHL} gene in over 90\% of cases. This genetic alteration leads to the stabilization of hypoxia inducible factor alpha  (HIF-$\alpha$), promoting angiogenesis and metabolic reprogramming. 

Despite initial curative surgical intervention, recurrence remains a major clinical concern in ccRCC. Approximately one-third of patients eventually develop regional or distant metastatic disease, and outcomes remain poor in advanced stages, with a five-year survival rate of only 13\% for those presenting with distant metastasis  \cite{kase2023clear}. Risk stratification tools like the TNM system and SSIGN score, which incorporate tumor stage, size, grade, and necrosis, remain clinically useful. However, as the SSIGN score was developed in an earlier treatment era, it does not fully capture the molecular heterogeneity and evolving biology of ccRCC that drive recurrence \cite{chong2025establishing}.
\subsection{Problem formulation}
The main challenge in predicting cancer recurrence is the inability of existing models to effectively integrate and analyze multiple heterogeneous data types. Traditional unimodal approaches rely on a single data modality (e.g., imaging, clinical, or molecular data), which limits prediction accuracy. These models also fail to capture complex, nonlinear relationships between biological, imaging, and clinical factors. Furthermore, the lack of personalized models tailored to individual patient characteristics results in less accurate predictions.

Let \( X_{\text{im}} \), \( X_{\text{cl}} \), and \( X_{\text{mo}} \) represent the imaging, clinical, and molecular data modalities, respectively. The recurrence prediction task can be modeled as:

\[
\hat{y} = f_{\text{model}}\left( X_{\text{im}}, X_{\text{cl}}, X_{\text{mo}} \right)
\]

Where:
\begin{itemize}
    \item \( \hat{y} \) is the predicted cancer recurrence outcome.
    \item \( f_{\text{model}} \) is the prediction model that integrates the different data modalities.
\end{itemize}

This problem can be further defined in terms of the integration of heterogeneous data sources, with the model needing to capture the complex relationships between multimodal features. For each modality, features \( \mathcal{F}_{\text{im}} \), \( \mathcal{F}_{\text{cl}} \), and \( \mathcal{F}_{\text{mo}} \) are extracted from \( X_{\text{im}} \), \( X_{\text{cl}} \), and \( X_{\text{mo}} \) respectively. These features are then combined to create a unified feature space \( \mathcal{F} \):

\[
\mathcal{F} = \mathcal{F}_{\text{im}} \oplus \mathcal{F}_{\text{cl}} \oplus \mathcal{F}_{\text{mo}}
\]

Where \( \oplus \) denotes a fusion operation (early or late fusion) that combines features from different modalities. The final prediction is:

\[
\hat{y} = g\left( \mathcal{F} \right)
\]

Where \( g(\mathcal{F}) \) represents the function that maps the combined feature space \( \mathcal{F} \) to the prediction outcome.

The challenges that arise during the modeling process are:
\begin{itemize}
    \item Effectively combining heterogeneous data types \( X_{\text{im}}, X_{\text{cl}}, X_{\text{mo}} \) to form a unified and meaningful feature set.
    \item Accurately capturing the nonlinear relationships between features from different modalities.
    \item Ensuring the prediction model is interpretable in clinical contexts, providing meaningful insights that go beyond a "black box" approach.
\end{itemize}
This work contributes to the field of cancer recurrence prediction by addressing the limitations of unimodal approaches through a novel multimodal integration framework. The key contributions include:

\begin{itemize}
    \item \textbf{Development of a Multimodal Framework:} The study proposes an integration framework that incorporates multiple data types:
    \begin{itemize}
        \item \textbf{CT/MRI Imaging:} Radiomic features are extracted to provide quantitative insights from medical imaging.
        \item \textbf{Histopathology:} The study leverages Vision Transformers and CNNs, along with CLAM, a weakly supervised method, to process whole-slide images (WSIs) for classification and region localization, eliminating the need for manual annotations.
        \item \textbf{EHRs \& Genomics:} MLPs (Multi-Layer Perceptrons) architectures are used to process structured clinical and genomic data, providing robust predictive modeling.
    \end{itemize}
    \item \textbf{Tailored Methodologies for Each Modality:} The study designs modality-specific methods to extract features from each data type, optimizing the utility of each data source:
    \begin{itemize}
        \item Radiomics-based feature extraction from CT/MRI imaging.
        \item CLAM-based classification and attention-guided region localization for WSIs.
        \item MLPs (Multi-Layer Perceptrons) used for feature extraction from EHR and genomic data.
    \end{itemize}
    \item \textbf{Comparative Analysis of Fusion Strategies:} The study evaluates multiple multimodal fusion strategies, including early/data-level fusion and late/decision-level fusion, to determine the most effective approach for improving predictive performance and optimizing integration across modalities.
\end{itemize}

\section{Related work}\label{sec2}
This section reviews recent developments in AI applied to cancer recurrence prediction. It highlights advances in unimodal and multimodal frameworks across imaging, genomics, and clinical data, with a focus on models integrating multiple modalities to overcome current limitations.
\subsection{Machine Learning for Cancer Recurrence Prediction}
DL techniques have revolutionized the field of biomedical data analysis, offering remarkable capabilities in predicting cancer recurrence. The pioneering work \cite{khan2025multi, khan2025detection} laid the groundwork for these innovations, introducing neural networks that have since become fundamental tools in medical diagnostics. The application of deep learning methods in cancer recurrence prediction leverages large datasets from medical imaging \cite{khan2025robust}, histopathological slides \cite{khan2024glnet}, and patient clinical data. These methods are particularly effective in extracting hidden patterns and subtle biomarkers that traditional methods might overlook, enabling more accurate and early detection of cancer recurrence.

\bmhead{Radiomics-Based Recurrence Prediction} 
Radiomics has emerged as a critical tool for quantifying tumor phenotypes from medical imaging modalities, such as CT and MRI. By extracting high-dimensional data from medical images, radiomics enables machine learning models to capture fine-grained imaging biomarkers associated with cancer recurrence. These biomarkers are often imperceptible to the human eye, yet they provide crucial insights into the tumor's characteristics and behavior.

In the context of hepatocellular carcinoma (HCC), Iseke et al. \cite{Iseke2023} developed a pipeline combining CNNs and XGBoost, integrating MRI features with clinical data to predict recurrence with an AUC of 0.76. This approach highlighted the power of combining imaging features with structured patient data to improve predictive accuracy. Similarly, Wang et al. \cite{Wang2023} applied deep learning on enhanced CT scans to predict recurrence in bladder cancer, achieving an AUC of 0.889, underscoring the potential of advanced imaging techniques in recurrence prediction.

For prostate cancer, Gu et al. \cite{Gu2023} developed NAFNet, a deep neural network trained on MRIs, which outperformed traditional models such as ResNet-50, achieving an impressive AUC of 0.915. This study emphasizes the ability of deep networks to capture intricate features from MRI scans, which can enhance the prediction of cancer recurrence. Additionally, Cepeda et al. \cite{Cepeda2023} addressed glioblastoma recurrence using voxel-based MRI radiomics and classifiers such as CatBoost and XGBoost, achieving an AUC of 0.81, further demonstrating the efficacy of radiomics in neuro-oncology.

\bmhead{Multimodal Fusion Approaches}
Despite the significant advances in unimodal data analysis, single-modal approaches often face limitations in capturing the full complexity of cancer recurrence. As a result, multimodal fusion strategies have gained increasing attention in recent years. By combining data from multiple sources, such as imaging, genomics, and clinical records, multimodal models can better address the diverse factors influencing recurrence and improve prediction accuracy.

Subramanian et al. \cite{subramanian2020multimodal} pioneered a multimodal approach for lung cancer recurrence prediction by combining imaging and genomics data. Their model demonstrated improved accuracy over isolated modalities, highlighting the complementary nature of these data sources. In a similar vein, Ren et al. \cite{Ren2023} combined MRI and clinical features using classifiers such as SVM and KNN, achieving AUCs of 0.965 and 0.955, respectively. This work focused on differentiating true glioma recurrence from treatment effects, a challenging task where multimodal data fusion provides significant advantages.

Qiu et al. \cite{Qiu2022} integrated H\&E histology images with molecular data for microsatellite instability classification in colorectal cancer, achieving an AUC of 0.952. This study highlights the value of combining traditional histopathological images with molecular data to enhance diagnostic accuracy. Similarly, Alinia et al. \cite{Alinia2024} used gradient boosting algorithms to predict recurrence in colorectal cancer, achieving an AUC of 0.964. The integration of multiple data types, including imaging and molecular features, has proven to be a key factor in improving prediction performance in cancer recurrence.

Further advancing multimodal approaches, Fu et al. \cite{Fu2023} proposed a deep multimodal graph-based model (DMGN), which combined multiplexed images and clinical variables to predict survival outcomes. By leveraging graph structures for data fusion, the model effectively captured complex relationships between various data modalities, improving the accuracy of survival predictions. The use of graph-based models is an innovative step in multimodal data integration, offering a flexible framework for handling diverse data types and enhancing model interpretability.

\subsection{Large Scale Multimodal Studies and Real-World Validation}
Large cohort, multi center datasets enhance generalizability. Noman et al. \cite{noman2025leveraging} merged METABRIC, MSK, Duke, and SEER data (n = 272,252) to predict breast cancer recurrence using survival analysis and ML models. Their best model (LightGBM) achieved AUC = 0.92, with external validation on Egyptian patients (84\% accuracy). Bone metastasis predictions were most reliable (AUC = 0.74), while brain/liver/lung differentiation remained difficult. Chen et al. \cite{chen2024multimodal} proposed a multimodal ensemble model (MMEM) for ccRCC prognosis by fusing WSI (UNI model), genomics, miRNA, methylation, and clinical data. Their method outperformed single modality models (C-index: 0.820 for OS, 0.833 for DFS). Challenges included visual interpretability and external validation.

Mahootiha et al. \cite{mahootiha2024multimodal} developed a CT+clinical multimodal deep learning model for RCC survival. A 3D CNN extracted radiomics, while clinical features were selected via random forests. Their model achieved a C-index of 0.84, supporting the clinical utility of radiomic-clinical fusion. Paverd et al. \cite{paverd2024radiology} categorized multimodal AI integration into three strategies: fusion, translation, and aggregation. They emphasized the value of 3D radiology for spatial insights and endorsed transformers and MIL as key tools for integrating radiology and molecular modalities. Alignment and heterogeneity challenges remain barriers to clinical adoption.

Digital pathology has advanced WSI-based prediction. Shi et al. \cite{Shi2023} trained CNNs on H\&E slides from the Carolina Breast Cancer Study for early recurrence prediction. Goyal et al. \cite{Goyal2024} used a multi-model approach integrating WSIs with clinicopathologic data, achieving state-of-the-art performance for breast cancer recurrence classification. Cross-modal transformers and privacy-preserving architectures have emerged. Goyal et al. \cite{Goyal2024} introduced a cross-modal transformer capturing spatial WSI features with clinical correlations.

\section{Proposed Research: MeD-3D: A Multimodal Fusion Framework for ccRCC}\label{sec3}
This section presents the methodological framework developed for cancer recurrence prediction in patients diagnosed with ccRCC. The proposed approach adopts a multimodal DL paradigm, integrating heterogeneous biomedical data sources to enhance predictive robustness and clinical applicability.

The overall methodology follows a multimodal DL framework designed to predict cancer recurrence in patients with ccRCC. As illustrated in Fig~\ref{fig:workflow}, the pipeline integrates clinical data, CT/MRI scans, and digital pathology WSI to extract modality specific features. Each data stream undergoes preprocessing, exploratory analysis, and model training using domain adapted architectures: a MLP for clinical/genomic data, a Med3D-based model for radiology data, and CLAM (Clustering-constrained Attention MIL) for histopathology slides.

Feature vectors from each modality are integrated using both early and late fusion strategies to enhance the robustness of recurrence prediction, particularly in scenarios with missing or incomplete data. This flexible multimodal approach leverages the complementary strengths of heterogeneous biomedical data sources clinical, radiological, and pathological to improve predictive performance, generalization, and adaptability to real-world clinical conditions.
\begin{figure}[H]
    \centering
    \includegraphics[width=\textwidth]{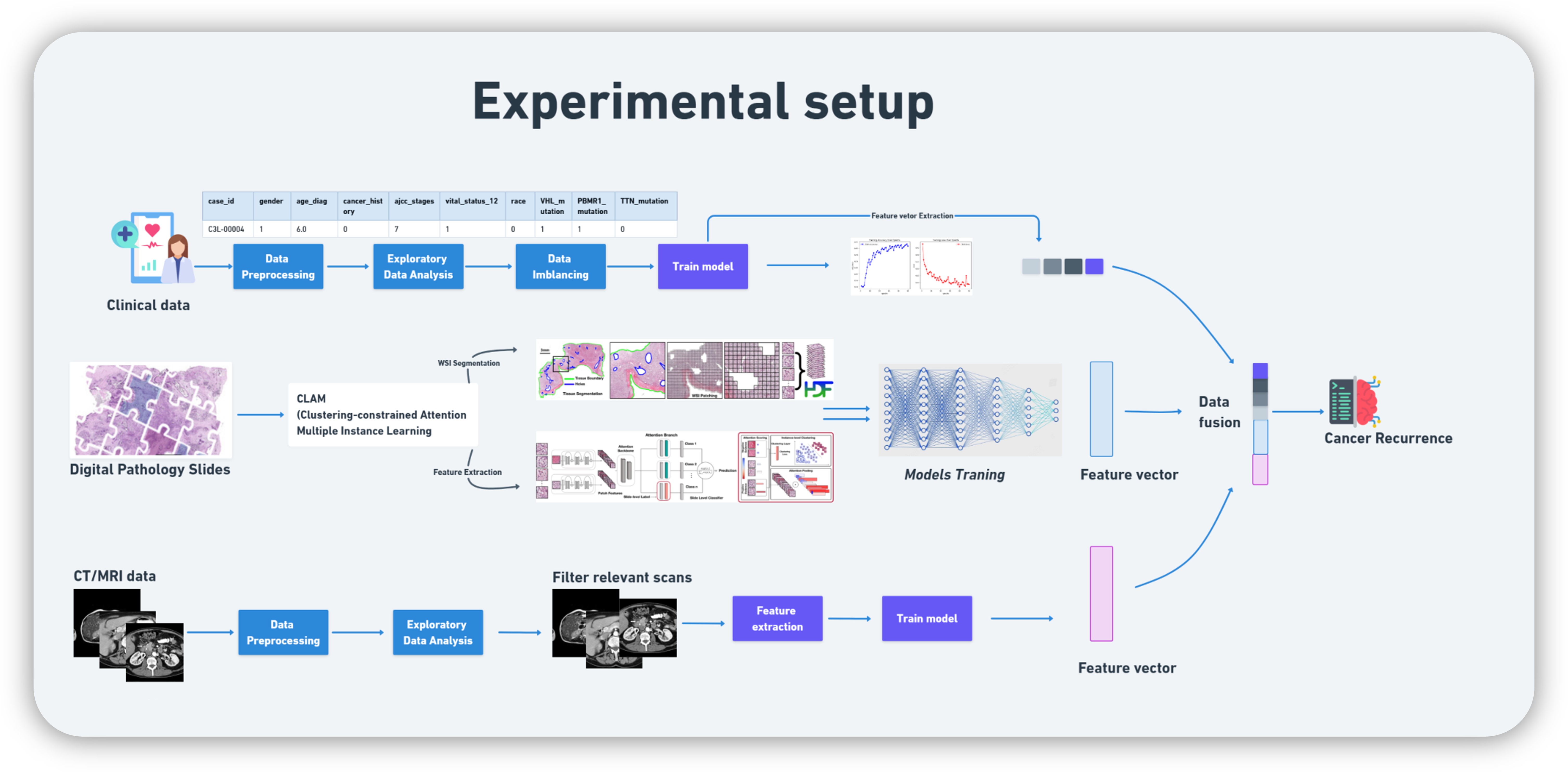}
    \caption{Workflow of the proposed multimodal cancer recurrence prediction pipeline.}
    \label{fig:workflow}
\end{figure}

\subsection{Dataset Collection}
\label{sec:Dataset_col}
\noindent\hspace{1.5em}The dataset was curated from two major public repositories: \href{https://www.cancerimagingarchive.net/collection/tcga-kirc/}{The Cancer Genome Atlas (TCGA)} and \href{https://www.cancerimagingarchive.net/collection/cptac-ccrcc/}{the Clinical Proteomic Tumor Analysis Consortium (CPTAC)}. It includes multimodal data from 618 patients diagnosed with clear cell renal cell carcinoma (ccRCC), encompassing structured electronic health records (EHR), histopathology whole-slide images (WSIs), and radiological scans (CT/MRI).

\subsection{Whole-Slide Imaging (WSI)}
\label{sec:wsi_p}
The histopathology modality leverages high-resolution Whole-Slide Images (WSIs) sourced from the TCGA-KIRC and CPTAC-CCRCC collections. This dataset comprises 2,573 H\&E-stained slides, representing 618 patients with varying numbers of slides per case. As illustrated in Figure~\ref{fig:workflow}, these gigapixel-sized images capture complex tissue morphology at a microscopic level, presenting a significant data processing challenge. To manage this, our proposed pipeline is based on the Clustering-constrained Attention Multiple Instance Learning (CLAM) framework. The pipeline first segments relevant tissue regions from the slide's background and then tiles these regions into thousands of smaller, manageable 256x256 pixel patches. Subsequently, a pretrained deep learning encoder, such as ResNet50, converts these patches into high-dimensional feature embeddings. To derive a single patient-level representation for multimodal fusion, these patch-level features are aggregated using an attention-based mechanism that identifies and weighs the most prognostically relevant regions. This process yields a final feature vector that encapsulates the critical morphological patterns from the histopathology data.

\begin{figure}[H]
    \centering
    \includegraphics[width=0.8\textwidth]{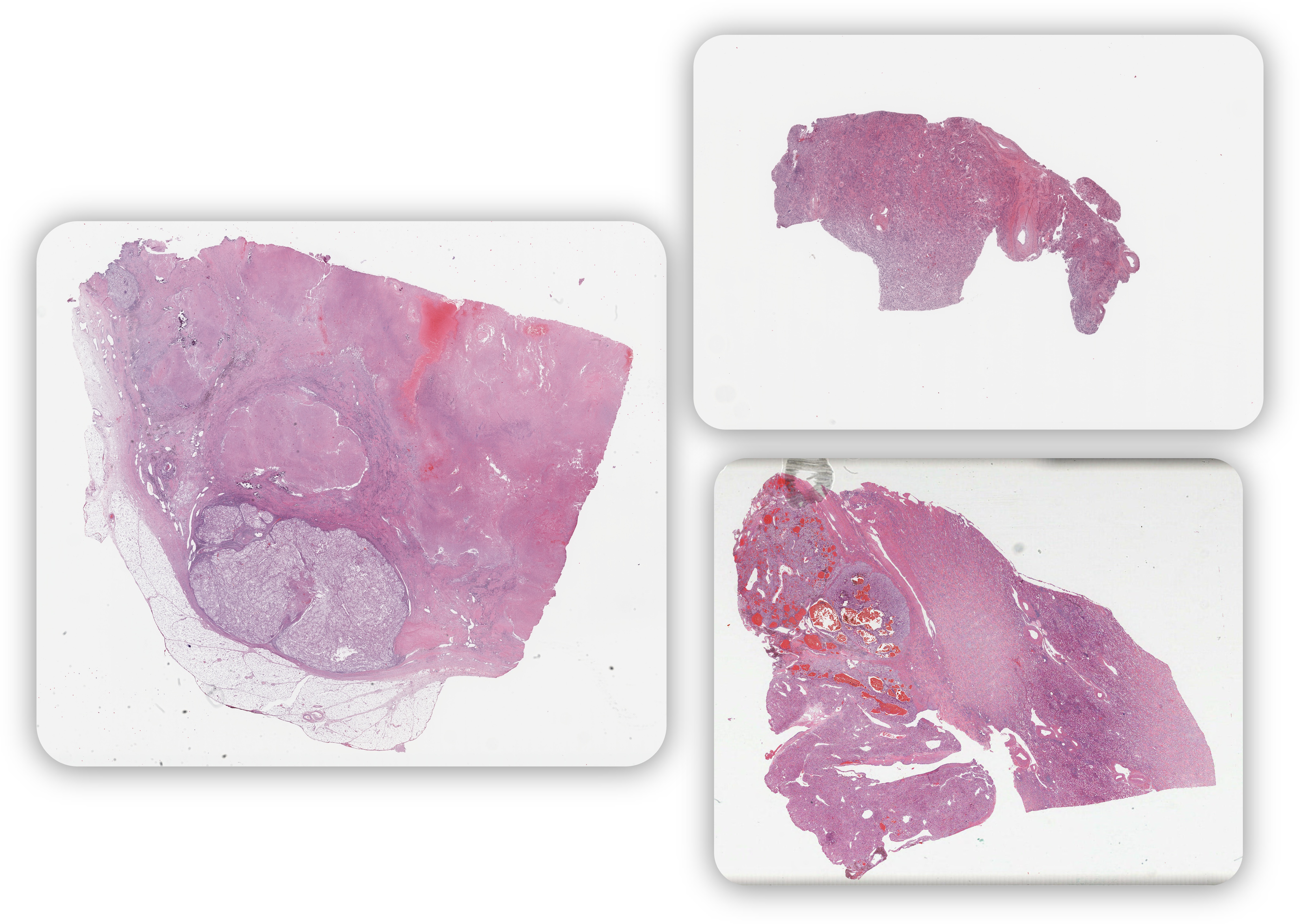}
    \caption{Samples of Whole sides images (WSI)}
    \label{fig:workflow}
\end{figure}

\subsection{Electronic Health Records (EHR)}
\label{sec:ehr_p}
The EHR modality pipeline begins with clinical and genomic data obtained from TCGA and CPTAC metadata. This structured dataset covers essential patient information, including demographics (\texttt{age\_diag}, \texttt{gender}), tumor staging according to AJCC criteria, and binary indicators for key gene mutations. Table~\ref{tab:ehr_sample} provides a representative snapshot of this data, illustrating the mix of numerical, categorical, and binary attributes.

To prepare this raw data for machine learning, our proposed pipeline employs a comprehensive preprocessing sequence. This involves handling missing values through median and mode imputation, encoding categorical features to preserve the clinical order of variables like tumor stage, and normalizing all numerical attributes to a uniform [0, 1] range. A critical component of the method is addressing the significant class imbalance in survival labels, which is managed by applying advanced oversampling techniques (SMOTE and ADASYN) to the training data. Once fully processed, the data is used to train a Multilayer Perceptron (MLP) for recurrence prediction. Finally, 128-dimensional feature embeddings are extracted from the MLP's final hidden layer, creating a compact and information-rich representation of the EHR modality for downstream multimodal fusion.


\begin{table}[htbp]
\centering
\caption{Clinical and Molecular Characteristics of Selected Cases}
\label{tab:ehr_sample}
\begin{tabular}{lccccccS[table-format=1.6]cc}
\toprule
\textbf{Case ID} & \textbf{Gender} & \textbf{Age} & \textbf{Grade} & \textbf{Stage} & \textbf{Vital} & \textbf{VHL} & \textbf{PBMR1} \\
 & \textbf{(M=1,F=0)} & & & & \textbf{Status} & \textbf{Mutation} & \textbf{Mutation} & \\
\midrule
C3L-01557 & 1 & 4.0 & 3 & III & 1 & 1 & 1 \\
C3N-01078 & 0 & N/A & 2 & N/A & 0 & 1 & 0 \\
C3N-00577 & 1 & 6.0 & 3 & IV & 1 & 0 & 1 & \\
TCGA-BP-4352 & 0 & 6.0 & 4 & IV & 0 & -1 & -1 \\
TCGA-A3-3307 & 1 & 5.0 & 3 & III & 1 & -1 & -1 \\
\bottomrule
\end{tabular}

\smallskip
\footnotesize
\textit{Note:} Stage is derived from AJCC pathological tumor stage. Vital status: 1 = Deceased, 0 = Living. Mutation status: 1 = Present, 0 = Absent, -1 = Unknown.
\end{table}

\subsection{Radiological Imaging (CT/MRI)}
\label{sec:ctmri_p}
\begin{figure}[H]
    \centering
    \begin{minipage}[t]{0.48\textwidth}
        \centering
        \includegraphics[width=\textwidth]{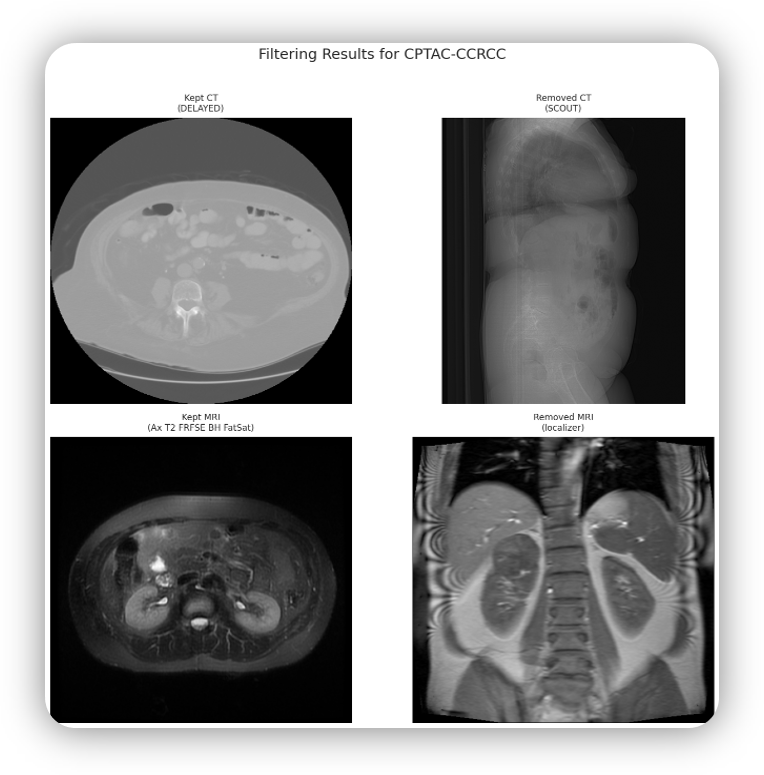}
        \begin{center}
     \textbf{(a)  CPTAC-CCRCC}    
     \end{center}
    \end{minipage}
    \hfill
    \begin{minipage}[t]{0.48\textwidth}
        \centering
        \includegraphics[width=\textwidth]{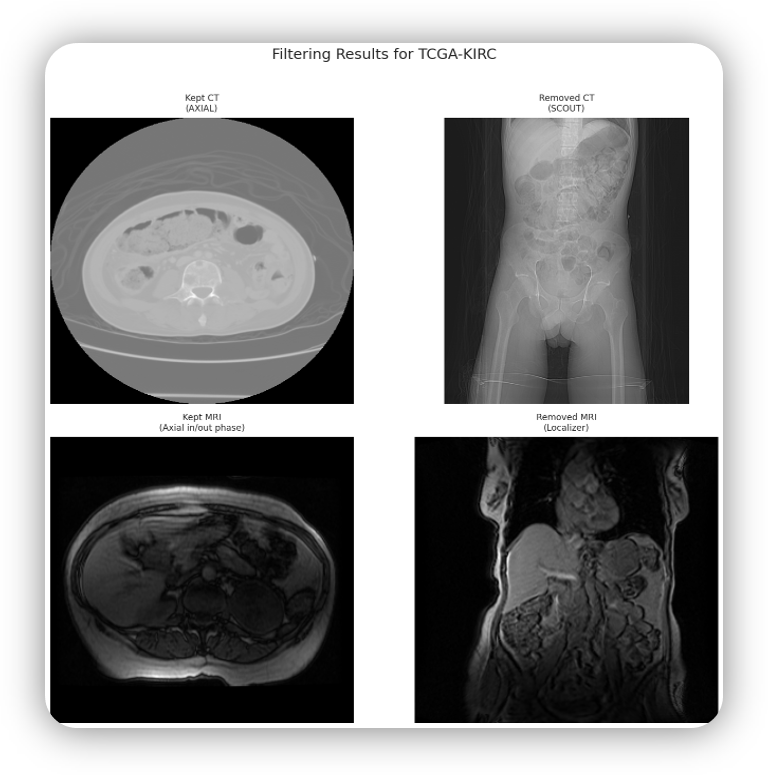}
               \begin{center}
     \textbf{(b)  TCGA-KIRC}    
     \end{center}
    \end{minipage}
    \caption{Illustration of filtering (a) CPTAC-CCRCC and (b) TCGA-KIRC cohorts.}
    \label{fig:filtering_process} 
\end{figure}

The radiological modality pipeline is built upon CT and MRI scans sourced from the TCGA-KIRC and CPTAC-CCRCC cohorts. The initial dataset comprised a large and heterogeneous collection of 3,464 scans (2,650 from TCGA and 814 from CPTAC). The first crucial step of our pipeline is a robust filtering process designed to isolate diagnostically relevant scans. As illustrated in Figure~\ref{fig:filtering_process}, we apply a keyword-based strategy to the \texttt{SeriesDescription} of each scan to retain only high-quality, axial-plane, post-contrast series. This systematic curation reduced the dataset to 907 diagnostic volumes, ensuring a consistent and clinically relevant cohort for analysis.

\begin{figure}[H]
    \centering
    \includegraphics[width=\textwidth]{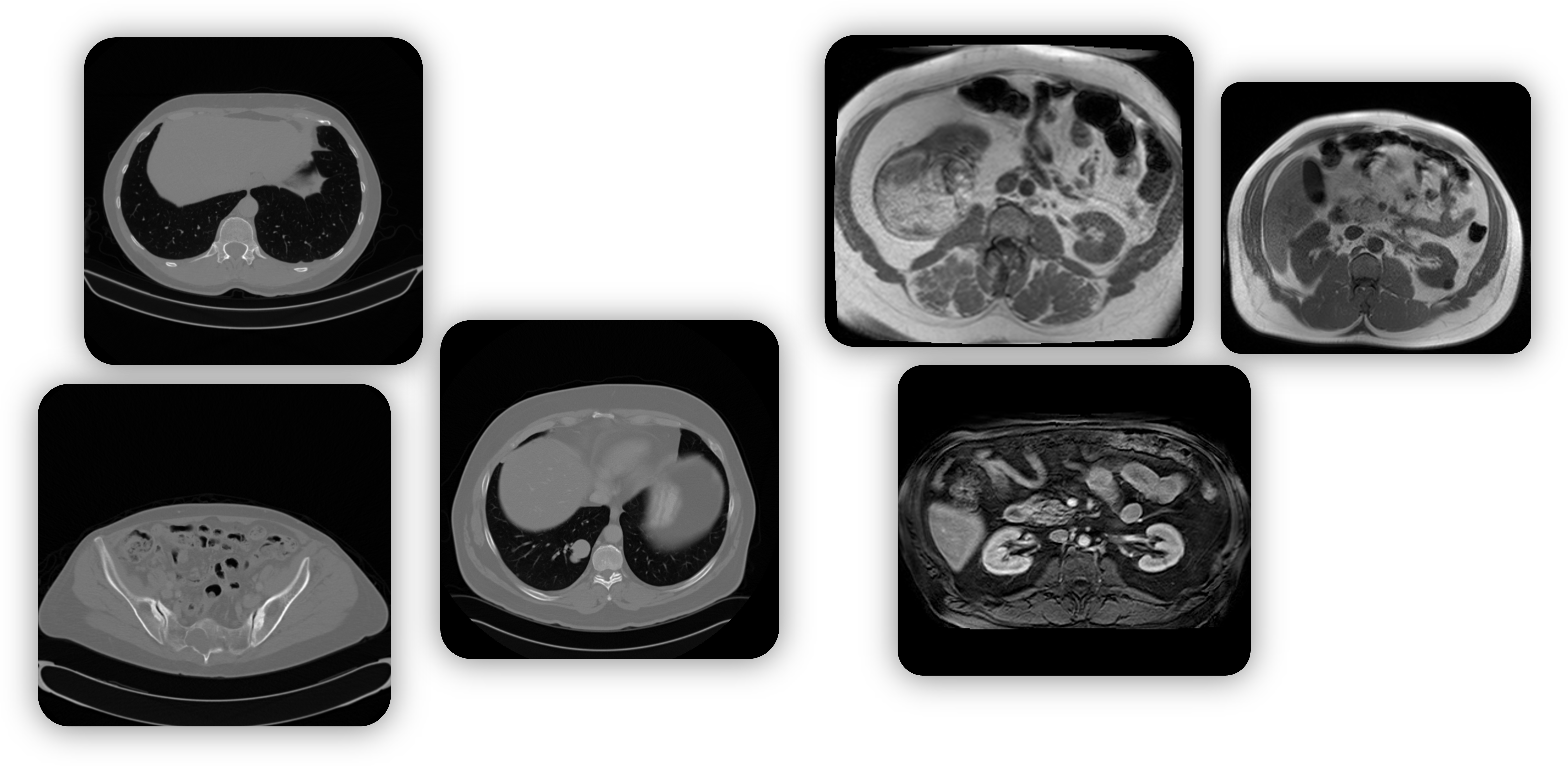}
    \caption{Samples of scans of CT and MRI}
    \label{fig:curated_scans_samples} 
\end{figure}

Examples of the resulting curated scans, which served as the input for feature extraction, are shown in Figure~\ref{fig:curated_scans_samples}. Each of these 3D volumes was then preprocessed through spatial resampling to a uniform resolution of $448 \times 448 \times 56$ and standardized intensity normalization. To extract powerful prognostic features, we employ a pretrained 3D ResNet-18 model from the MedicalNet framework. This model processes each scan to generate a 512-dimensional feature embedding. Finally, to create a single patient-level representation for multimodal fusion, the embeddings from all of a patient's eligible scans are aggregated, yielding the final feature set for the radiological modality.

\subsection{Multimodal Fusion Strategies}
\label{sec:mm_fusion}
To fully leverage the complementary strengths of the EHR, WSI, and CT/MRI modalities, we propose and evaluate both early and late fusion strategies. The foundation for these strategies is a harmonized, patient-level feature table constructed by merging pre-computed embeddings from each data stream. As illustrated in Figures \ref{fig:ehr_features}, \ref{fig:wsi_features}, and \ref{fig:ctmri_features}, these embeddings are high-dimensional vectors that represent the salient information from each modality: a 64-dimensional vector for EHR, a 1024-dimensional vector for WSI, and a 512-dimensional vector for CT/MRI. These unified feature sets serve as the input for the fusion models described below.

\bmhead{Fusion Dataset Construction.}

Feature embeddings were precomputed separately for each modality and stored as structured tabular files:

\begin{itemize}
    \item \textbf{EHR:} A 64-dimensional feature vector was extracted from the fc3 layer of the trained MLP classifier. Each row corresponds to one patient and was stored in \texttt{clinical\_features.csv}. These embeddings represent latent clinical and genomic patterns useful for prediction.

    \begin{figure}[h!]
        \centering
        \includegraphics[width=0.9\textwidth]{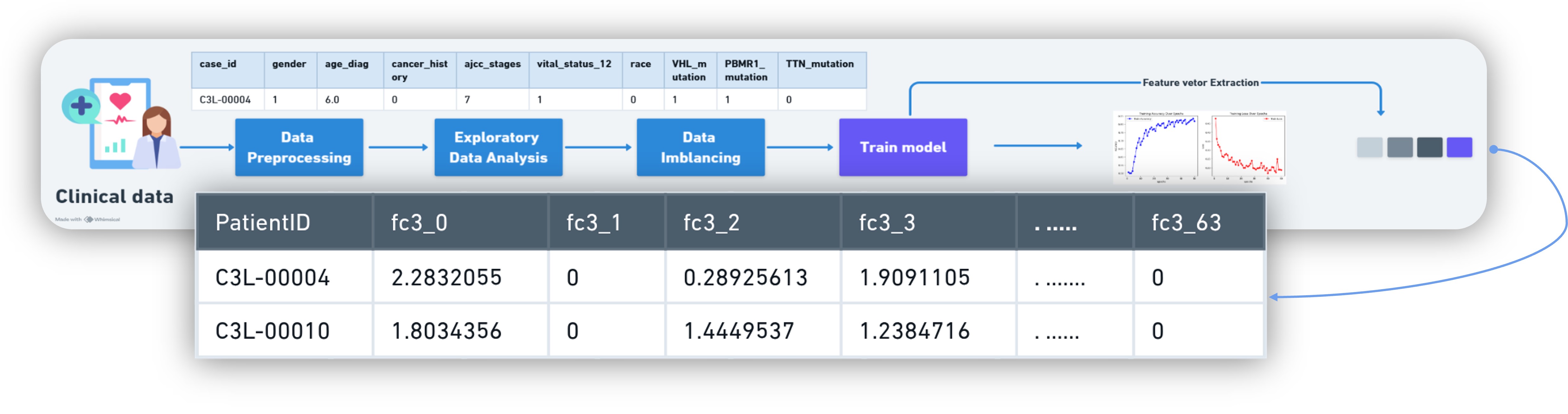}
        \caption{EHR modality: Extracted 64-dimensional fc3 embeddings for each patient after training the MLP classifier.}
        \label{fig:ehr_features}
    \end{figure}
    \item \textbf{WSI:} Using the CLAM framework, slide-level patch embeddings were aggregated via attention pooling to generate a 1024-dimensional vector per patient. These were saved in \texttt{wsi\_features.csv} and capture spatial and morphological patterns across WSIs.
    \begin{figure}[h!]
        \centering
        \includegraphics[width=0.9\textwidth]{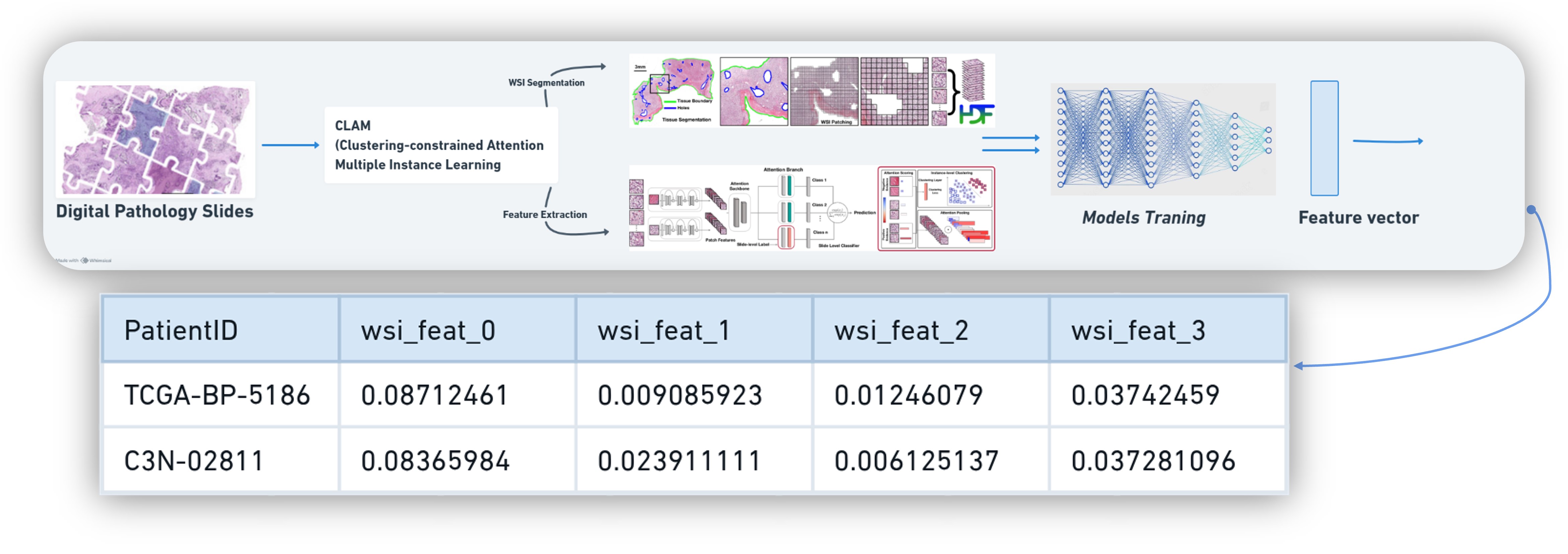}
        \caption{WSI modality: Aggregated attention-based patch-level features using the CLAM model. Each vector represents a single patient histopathological profile.}
        \label{fig:wsi_features}
    \end{figure}

    \item \textbf{CT/MRI:} Medical scans were processed using a Multiple Instance Learning pipeline with a 3D-ResNet18 backbone from MeD3D. 
    For each patient, features from the most informative scan were globally pooled into a 512-dimensional vector and saved in 
    \texttt{ct\_mri\_features.csv}.

    \begin{figure}[h!]
        \centering
        \includegraphics[width=0.9\textwidth]{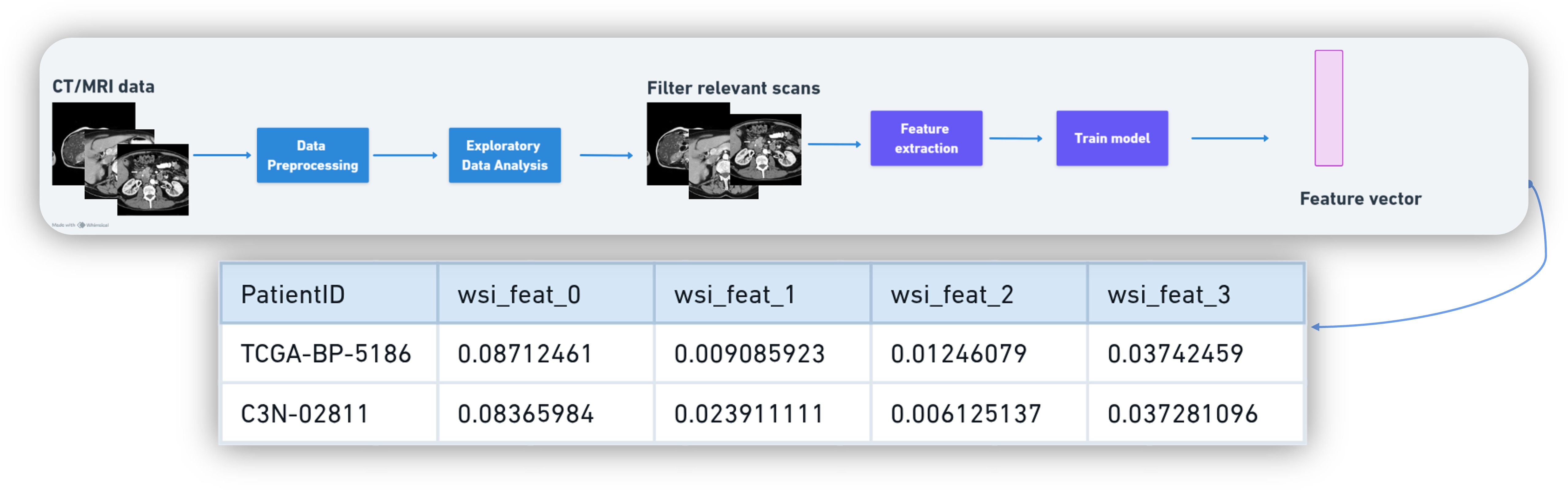}
        \caption{CT/MRI modality: Patient-level scan embeddings extracted from the MedicalNet pipeline using a 3D ResNet18 model.}
        \label{fig:ctmri_features}
    \end{figure}
\end{itemize}

\bmhead{Early Fusion: Feature-Level Integration}
In the early fusion approach, we combine the raw feature embeddings from all modalities before they are fed into a predictive model. This allows the model to learn complex, cross-modal interactions directly from the integrated feature space. We investigate two primary techniques:
\begin{itemize}
    \item \textbf{Concatenation:} The feature vectors from EHR, WSI, and CT/MRI are concatenated into a single, high-dimensional vector. This combined vector is then passed to a unified MLP classifier for final prediction.
    \item \textbf{Mean Pooling:} The feature vectors are element-wise averaged to create a single mean embedding, which is then used as input for the final classifier.
\end{itemize}

\bmhead{Late Fusion: Decision-Level Integration}
In the late fusion approach, each modality is first used to train an independent, specialized model. Each model produces a separate prediction probability for a given patient. These individual predictions are then combined at the decision level using two methods:
\begin{itemize}
    \item \textbf{Weighted Sum:} The final prediction is calculated as a weighted average of the individual probabilities, where each weight is proportional to the unimodal model's balanced accuracy on a validation set. This prioritizes the predictions from more reliable modalities.
    \item \textbf{Learned Weights:} A lightweight fusion network is trained to learn the optimal weights for combining the modality-specific predictions, allowing for a more dynamic and data-driven integration.
\end{itemize}

This strategy enabled the model to learn interactions across modalities and improved generalization when complete data was available.

\section{Results and implementation}\label{sec13}
This section presents the experimental setup, datasets used, evaluation metrics, and results obtained from training and evaluating models for prediction using  WSI, radiological (CT/MRI), and clinical (EHR) data. The experiments benchmark different modeling and balancing strategies, including the use of CLAM for WSI, and MLPs for EHR.
\subsection{Experimental settings}
The experimental setup and implementation were carried out using the following tools and software environments:

\begin{itemize}
    \item \textbf{Software Stack}: Python 3.10 with PyTorch, NumPy, pandas, imbalanced-learn, and matplotlib.
    \item \textbf{Deep Learning Frameworks}: CLAM (weakly-supervised attention MIL) for WSIs; MedicalNet (3D ResNet) for CT/MRI; MLP for EHR.
    \item \textbf{Infrastructure}: Experiments were conducted on a Linux-based high-performance computing cluster with NVIDIA A100 GPUs.
\end{itemize}



\section{Reporting and Compliance with TRIPOD Checklist}
\label{sec:tripod}

Table~\ref{tab:tripod_checklist} presents the TRIPOD (Transparent Reporting of a multivariable prediction model for Individual Prognosis Or Diagnosis) checklist, highlighting reporting items and their corresponding sections in this study. This ensures transparency and alignment with established standards for developing and evaluating prediction models in medical research.

\begin{table}[htbp]
\centering
\caption{TRIPOD Checklist Reporting}
\label{tab:tripod_checklist}
\begin{tabular}{p{2.8cm}p{7cm}p{3cm}p{1.2cm}}
\hline
\textbf{Section/Topic} & \textbf{Checklist Item} & \textbf{Reported on Page} & \textbf{Outcome} \\
\hline
\multicolumn{3}{c}{\textbf{Title and Abstract}} \\
\hline
Title & 1. Identify the study as developing and/or validating a multivariable prediction model & Title Page & Pass \\

Abstract & 2. Provide an abstract summarizing objectives, study design, results, and conclusions & Abstract & Pass\\
\hline
\multicolumn{3}{c}{\textbf{Introduction}} \\
\hline
Background & 3a. Explain the medical context & Sec.~\ref{sec3} & Pass \\
& 3b. Explain the prediction research context & Sec.~\ref{sec3} & Pass\\ 

Objectives & 4. Specify the objectives & Sec.~\ref{sec3} & Pass\\ 
\hline
\multicolumn{3}{c}{\textbf{Methods}} \\
\hline
Data Sources & 5. Describe the study design or data source & Sec.~\ref{sec:Dataset_col} & Pass\\

Whole-Slide Imaging (WSI) & 6. Histogram sample visualization & Sec.~\ref{sec:wsi_p} & Pass\\

Electrponic Health Record & 9. Patient records & Sec.~\ref{sec:ehr_p}& Pass \\

Radiological imaging & 10. Quantitative Radiomic Features & Sec.~\ref{sec:ctmri_p} & Pass\\

Sample Size & 11. Explain how sample size was determined & Sec.~\ref{sec:Dataset_col} & Pass\\

Analysis & 13. Describe modeling technique & Sec.~\ref{sec:mm_fusion} & Pass\\
& 14. Specify all measures of model performance & Sec.~\ref{sec:mm_fusion} & Pass\\
\hline
\multicolumn{3}{c}{\textbf{Results}} \\
\hline
Experiments on Clinical EHR Data & 15. Outcome on Clinical EHR Data & Sec.~\ref{ECE} & Pass\\

Radiological Imaging (CT/MRI) & 16. Outcome on Radiological Imaging (CT/MRI) & Sec.~\ref{sec:ct_mri_method} & Pass\\

Whole-Slide Imaging (WSI) & 17. Outcome on Whole-Slide Imaging (WSI) & Sec.~\ref{WS} & Pass\\

Multimodal Fusion Experiments & 18. Outcome on Multimodal Fusion Experiments & Sec.~\ref{MF} & Pass\\
\hline
\multicolumn{4}{c}{\textbf{Discussion}} \\
\hline
Limitations & 19. Acknowledge study limitations and the need for validation & Sec.~\ref{sec:future_work} & Pass\\

Interpretation & 20. Interpret comparative performance of fusion models & Sec.~\ref{sec:discussion} & Pass\\

Implications & 21. Discuss implications for prognostic modeling and clinical utility & Sec.~\ref{sec:discussion}, \ref{sec:future_work} & Pass\\
\hline

\multicolumn{4}{c}{\textbf{Conclusion}} \\
\hline
Conclusion & 22. Summarize key findings and future outlook & Sec.~\ref{sec:Con}, \ref{sec:future_work} & Pass \\
\hline
\multicolumn{4}{c}{\textbf{Other Information}} \\
\hline
Declarations & 23. Report funding, ethics, and competing interests & Sec.~\ref{sec:declar} & Pass \\

Abbreviations & 24. Define key terms and abbreviations used & Sec.~\ref{sec:abbrev} & Pass \\
\hline

\end{tabular}
\end{table}

The complete TRIPOD checklist is available as Supplementary Material 1. Our study complies with all applicable TRIPOD guidelines for transparent reporting of prediction model studies.

\subsection{Experiments on Clinical EHR Data}
\label{ECE}

\subsubsection{Data Preprocessing}
The raw clinical dataset consisted of 618 patient records and 20 features, including demographic, genetic, and pathological attributes. Prior to training, the dataset underwent a series of preprocessing steps to handle missing values, encode categorical variables, and scale numerical features. The goal was to ensure data consistency and suitability for input into a neural network model.

\begin{itemize}
    \item \textbf{Missing Value Imputation:} 
    Numerical columns such as \texttt{age\_diag} were imputed using the median value to reduce the influence of outliers. For categorical features, the most frequent value (mode) was used. Specifically, the column \texttt{cancer\_history}, which had substantial missingness, was imputed using the most frequent class. Additionally, special placeholder values (e.g., \texttt{-1} in \texttt{ajcc\_path\_tumor\_stage}) were treated as missing and appropriately replaced.

    \item \textbf{Categorical Feature Encoding:} 
    Ordinal encoding was applied to staging-related variables to preserve their inherent order:
    
    \begin{itemize}
        \item \verb|ajcc_path_tumor_stage|
        \item \verb|ajcc_path_tumor_pt|
        \item \verb|ajcc_path_nodes_pn|
        \item \verb|ajcc_clin_metastasis_cm|
        \item \verb|ajcc_path_metastasis_pm|
    \end{itemize}
    
    Binary variables such as \verb|gender| were one-hot encoded. The first category was dropped to avoid multicollinearity.

    \item \textbf{Feature Scaling:} 
    All numerical features were normalized using Min-Max scaling to bring them into the range [0, 1]. This step was essential to stabilize the training of the neural network and ensure uniform feature contributions. The scaled features included demographic variables (e.g., \texttt{age\_diag}) and all encoded tumor staging variables.

    \item \textbf{Feature Selection:} 
    Non-informative columns such as patient identifiers (\texttt{case\_id}) and data split indicators (\texttt{Split}) were removed from the feature set. The final input matrix $X$ was composed of all relevant clinical and genomic features, excluding the target variable \texttt{vital\_status\_12}, which was used as the binary class label $y$.
\end{itemize}

This comprehensive preprocessing pipeline ensured that the clinical data was clean, numerically stable, and ready for downstream modeling using machine learning classifiers.

\subsubsection{Class Balancing Methods}
To address the class imbalance in survival labels, we experimented with two oversampling strategies: \textbf{SMOTE} and \textbf{ADASYN}. Both methods were applied only to the training set to prevent information leakage. The aim was to improve the model’s ability to generalize to the minority class by ensuring that the classifier is exposed to a more balanced data distribution during training.

\bmhead{Synthetic Minority Over-sampling Technique}
SMOTE generates new synthetic instances of the minority class by interpolating between existing examples and their nearest neighbors in feature space.

\begin{figure}[h!]
    \centering
    \includegraphics[width=0.9\linewidth]{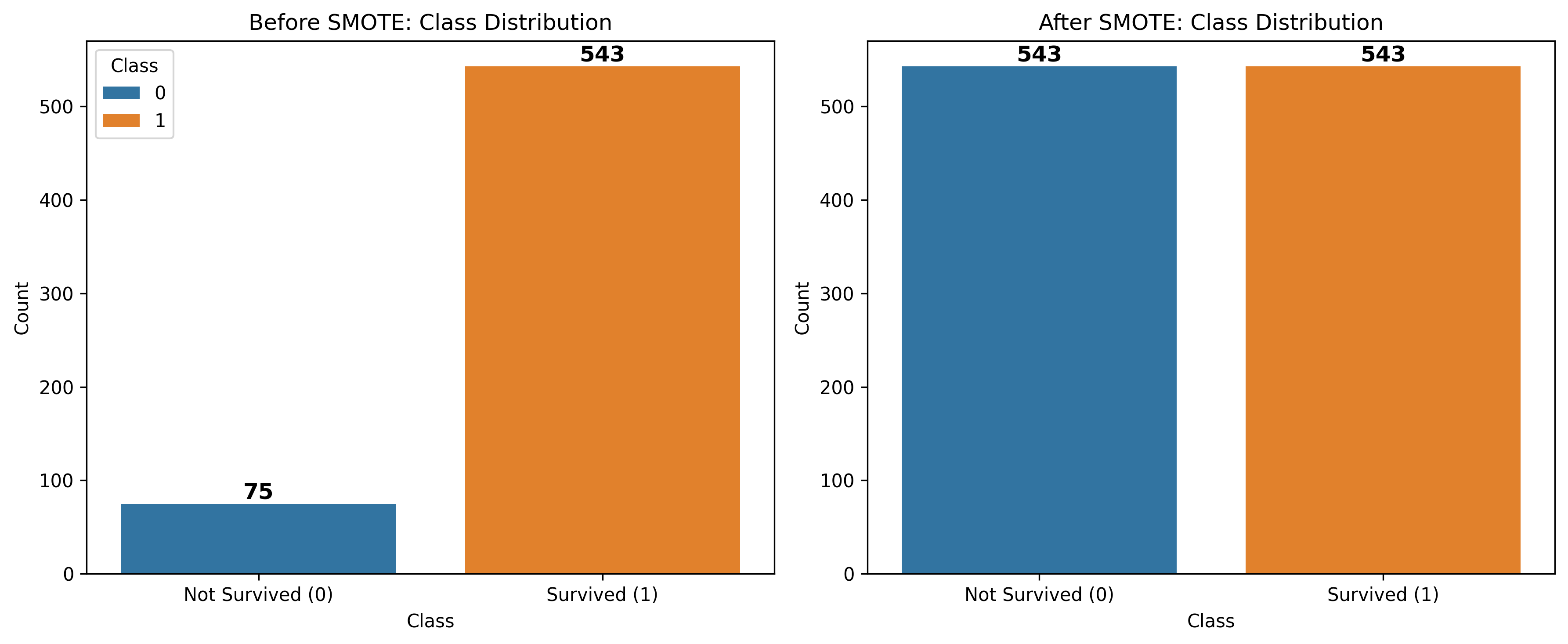}
    \caption{Class distribution before and after applying SMOTE.}
    \label{fig:smote_dist}
\end{figure}

As shown in Figure~\ref{fig:smote_dist}, SMOTE effectively balanced the class distribution by oversampling the minority class (Not Survived) to match the number of majority class samples. This helped mitigate the classifier’s tendency to be biased toward the dominant class during training.

\bmhead{Adaptive Synthetic Sampling}
ADASYN extends SMOTE by adaptively deciding where to generate synthetic samples. Instead of generating an equal number of synthetic points for all minority class samples, ADASYN focuses more on samples that are harder to learn  those surrounded by majority class points. This adaptive approach improves the model’s ability to generalize, especially near class boundaries.

\begin{figure}[H]
    \centering
    \includegraphics[width=0.9\linewidth]{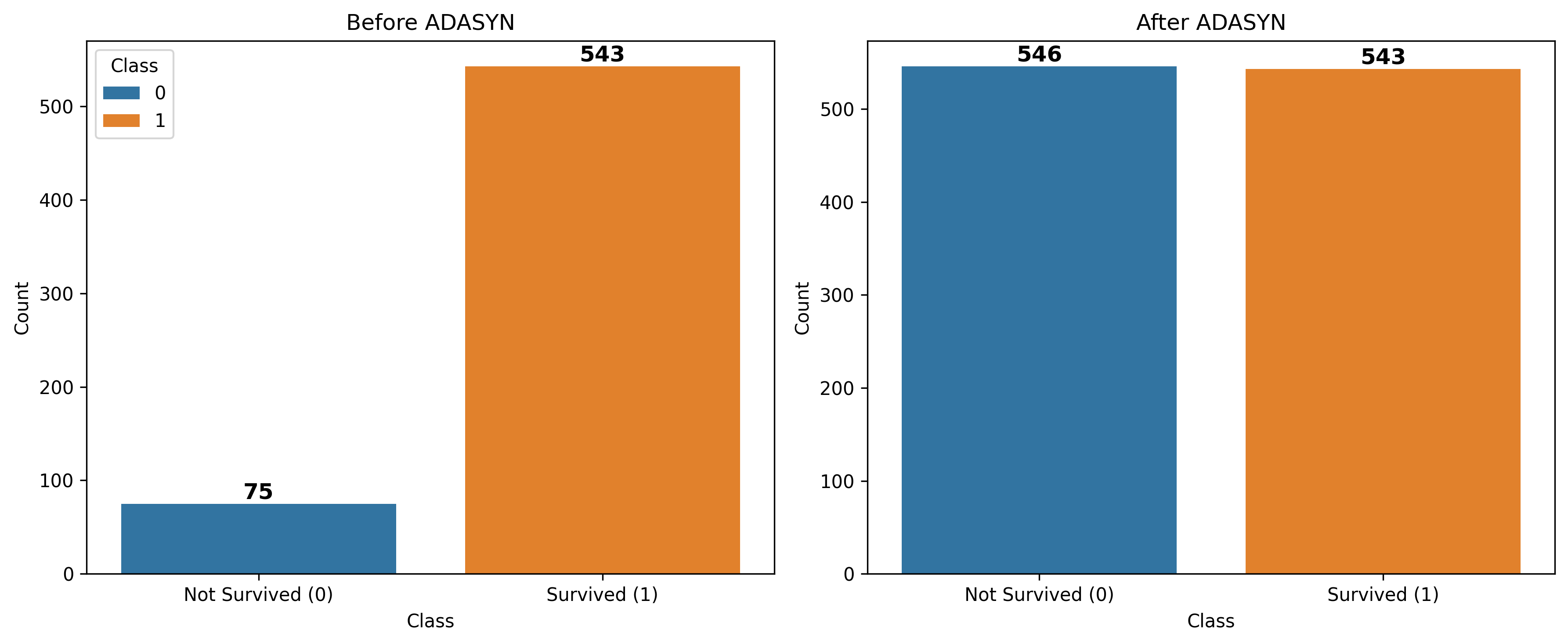}
    \caption{Class distribution before and after applying ADASYN.}
    \label{fig:adasyn_dist}
\end{figure}

As shown in Figure~\ref{fig:adasyn_dist}, ADASYN also rebalanced the dataset but did so adaptively, producing a slightly higher number of synthetic samples for the minority class. This targeted approach aims to improve performance on ambiguous or overlapping decision boundaries and may enhance model sensitivity to harder cases.

\subsubsection{Training Dynamics}

\bmhead{Model Architecture and Training}
A MLP classifier was designed to trained model on both the SMOTE  and ADASYN balanced datasets using the same architecture and hyperparameter:

\begin{itemize}
    \item \textbf{Layers:} [Input → 256 → 128 → 64 → Output]
    \item \textbf{Activation:} ReLU for hidden layers, Sigmoid for binary classification
    \item \textbf{Regularization:} Dropout (0.3) applied to first two hidden layers; Batch Normalization applied after each hidden layer
    \item \textbf{Loss Function:} Weighted CrossEntropyLoss to counter class imbalance
    \item \textbf{Optimizer:} Adam with weight decay and learning rate scheduler
    \item \textbf{Training Epochs:} 50
\end{itemize}

Training loss and accuracy were tracked for both SMOTE and ADASYN cases. The learning curves are shown below:

\begin{figure}[h!]
    \centering
    \includegraphics[width=0.9\linewidth]{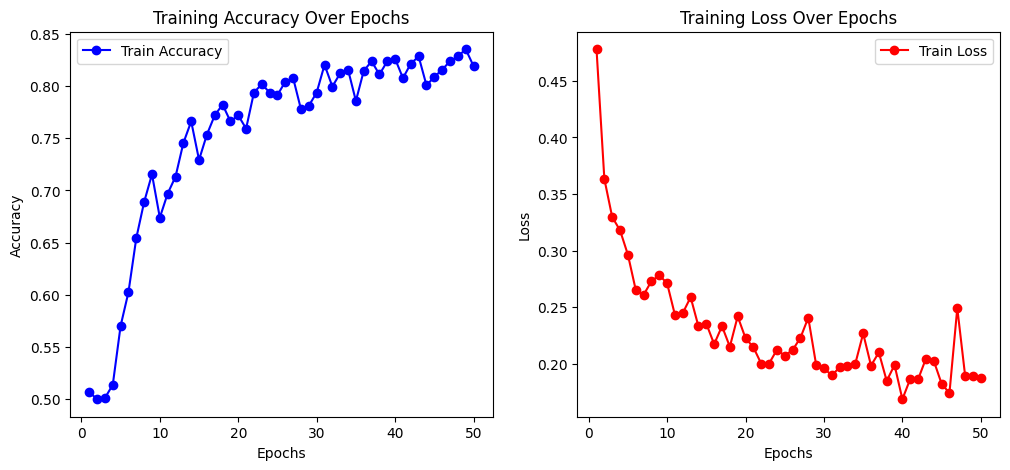}
    \caption{SMOTE: Training Accuracy and Loss over epochs.}
    \label{fig:train_curves_smote}
\end{figure}

\begin{figure}[H]
    \centering
    \includegraphics[width=0.9\linewidth]{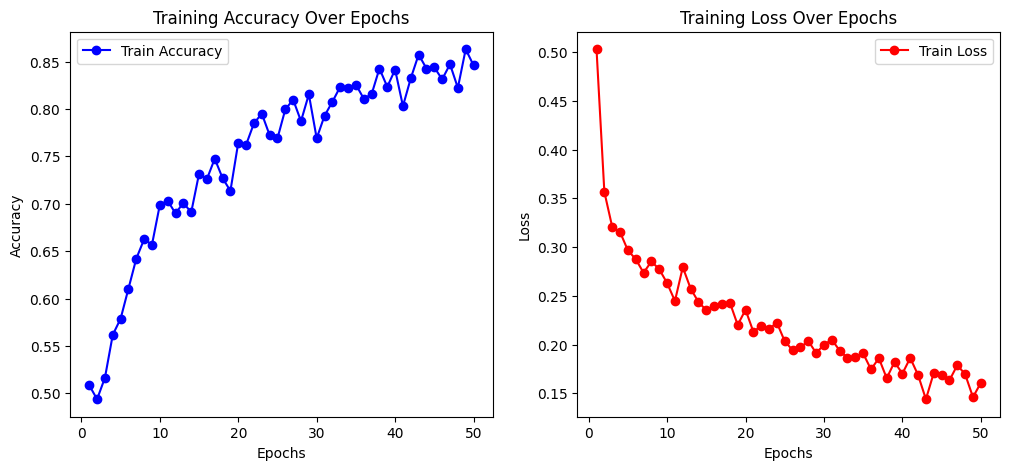}
    \caption{ADASYN: Training Accuracy and Loss over epochs.}
    \label{fig:train_curves_adasyn}
\end{figure}

\subsubsection{Evaluation Results}
\bmhead{SMOTE Results}
The model trained on the SMOTE balanced dataset yielded the following classification metrics:

The detailed classification metrics are presented in Table~\ref{tab:test_perf_smote_class} and Table~\ref{tab:test_perf_smote_overall}. As shown in the per-class report, the model achieved high recall (0.96) for the majority "Survived" class but a lower recall (0.77) for the minority "Not Survived" class. This indicates that while SMOTE helps, a slight bias towards the majority class persists. The overall test accuracy reached 86.70\%, providing a strong baseline performance.

\begin{table}[h!]
    \centering
    \caption{SMOTE: Classification Report on Test Set (Per-Class Metrics)}
    \begin{tabular}{lccc}
    \toprule
    Class & Precision & Recall & F1-score  \\
    \midrule
    Not Survived (0) & 0.95 & 0.77 & 0.85 \\
    Survived (1)     & 0.81 & 0.96 & 0.88  \\
    \bottomrule
    \end{tabular}
    
    \label{tab:test_perf_smote_class}
\end{table}

\begin{table}[h!]
    \centering
     \caption{SMOTE: Overall Test Set Metrics}
    \begin{tabular}{lc}
    \toprule
    Metric & Value \\
    \midrule
    Test Loss     & 0.1577 \\
    Test Accuracy & 0.8670 \\
    Precision & 0.8077 \\
    Recall    & 0.9633 \\
    F1 Score  & 0.8787 \\
    \bottomrule
    \end{tabular}
    \label{tab:test_perf_smote_overall}
\end{table}

\begin{figure}[h!]
    \centering
    \includegraphics[width=0.6\linewidth]{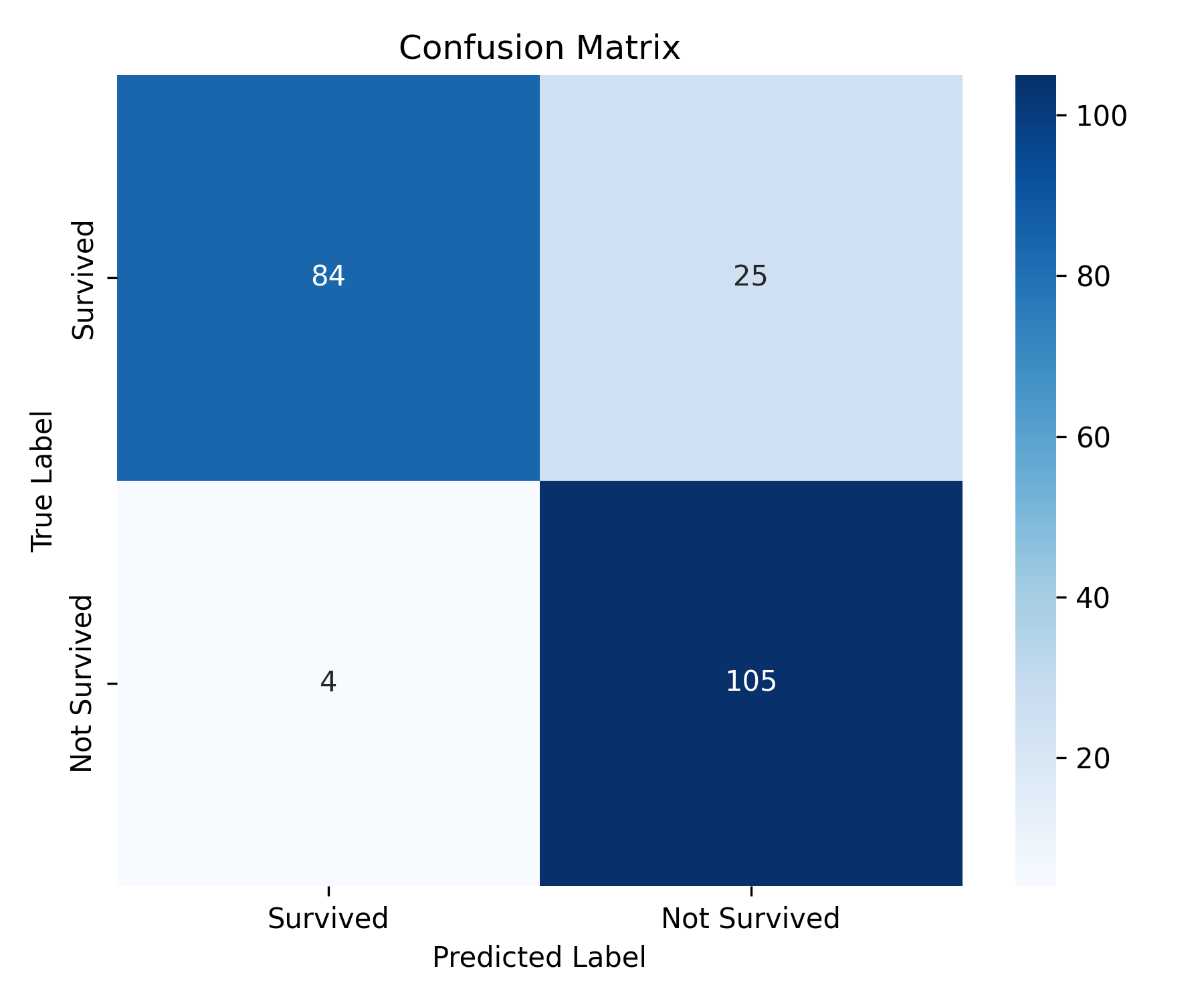}
    \caption{SMOTE: Confusion Matrix and ROC Curve}
    \label{fig:conf_roc_smote}
\end{figure}

\bmhead{ADASYN Results}
The model trained on the ADASYN balanced dataset showed slightly improved performance:

As detailed in Table~\ref{tab:test_perf_adasyn_class}, the ADASYN trained model showed a notable improvement in identifying the minority class, with recall for "Not Survived" increasing to 0.88. This balanced performance is reflected in the strong F1-scores for both classes (0.91 and 0.92). The overall metrics in Table~\ref{tab:test_perf_adasyn_overall} confirm this superiority, with the test accuracy rising to 91.74\% and a lower test loss of 0.1351, suggesting better generalization.

\begin{table}[h!]
    \centering
    \caption{ADASYN: Classification Report on Test Set (Per-Class Metrics)}
    \begin{tabular}{lcccc}
    \toprule
    Class & Precision & Recall & F1-score  \\
    \midrule
    Not Survived (0) & 0.95 & 0.88 & 0.91  \\
    Survived (1)     & 0.89 & 0.95 & 0.92  \\
    \bottomrule
    \end{tabular}
    \label{tab:test_perf_adasyn_class}
\end{table}

\begin{table}[h!]
    \centering
    \caption{ADASYN: Overall Test Set Metrics}
    \begin{tabular}{lc}
    \toprule
    Metric & Value \\
    \midrule
    Test Loss       & 0.1351 \\
    Test Accuracy   & 0.9174 \\
    Precision & 0.8889 \\
    Recall    & 0.9541 \\
    F1 Score  & 0.9204 \\
    \bottomrule
    \end{tabular}
    \label{tab:test_perf_adasyn_overall}
\end{table}

\begin{figure}[h!]
    \centering
    \includegraphics[width=0.6\linewidth]{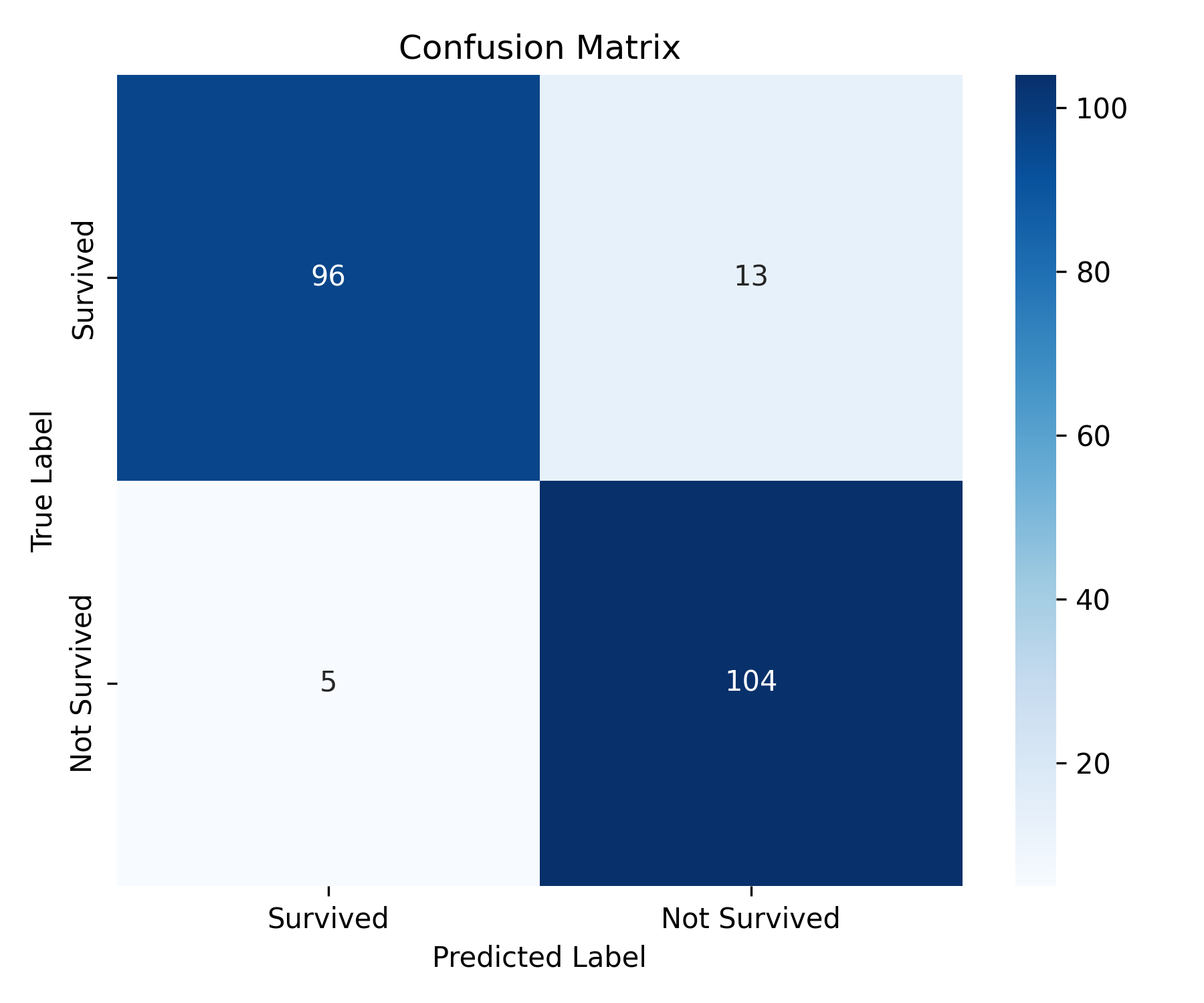}
    \caption{ADASYN: Confusion Matrix and ROC Curve}
    \label{fig:conf_roc_adasyn}
\end{figure}

\subsubsection{Comparison of Balancing Methods}
The table below summarizes the comparative performance between SMOTE and ADASYN on the test set:

Table~\ref{tab:compare_balancing} provides a direct comparison of the key performance metrics. The results clearly show that ADASYN outperformed SMOTE across all metrics. The most significant improvement was observed in the F1-score for the minority class ("Not Survived"), which increased from 0.85 to 0.91. This demonstrates the effectiveness of ADASYN's adaptive approach in generating more useful synthetic samples for hard-to-learn instances, leading to a more robust and clinically relevant model.

\begin{table}[h!]
    \centering
    \caption{Performance Comparison Between SMOTE and ADASYN}
    \begin{tabular}{lcc}
    \toprule
    Metric & SMOTE & ADASYN \\
    \midrule
    Accuracy & 0.8670 & 0.9174 \\
    F1-score (Class 0) & 0.85 & 0.91 \\
    F1-score (Class 1) & 0.88 & 0.92 \\
    \bottomrule
    \end{tabular}
    
    \label{tab:compare_balancing}
\end{table}

\bmhead{Outcome for EHR Modality}
Both SMOTE and ADASYN effectively addressed class imbalance and improved predictive performance. ADASYN showed slightly better generalization, particularly for the minority class, likely due to its focus on harder-to-learn samples. This makes it a strong candidate for class balancing in clinical prediction pipelines where sensitivity is critical.

\subsection{Radiological Imaging (CT/MRI)}
\label{sec:ct_mri_method}
This section describes the acquisition, processing, and modeling of radiological imaging data (CT and MRI) for predicting cancer recurrence in patients with clear cell Renal Cell Carcinoma (ccRCC). The comprehensive pipeline includes data acquisition, stringent filtering, 3D feature extraction using a pre-trained Convolutional Neural Network (CNN), and patient-level aggregation.

\subsubsection{Data Acquisition and Filtering}
\label{sec:ct_mri_data_filtering}
Radiological scans were retrieved from two public cohorts: \texttt{TCGA-KIRC} and \texttt{CPTAC-CCRCC}, via the \texttt{tcia-utils} API. An initial dataset of 2,650 scans from TCGA-KIRC and 814 scans from CPTAC-CCRCC (across CT and MRI modalities) was obtained.

A robust keyword  and regex based filtering strategy was applied to the \texttt{SeriesDescription} field to retain diagnostically relevant series and ensure consistency:
\begin{itemize}[leftmargin=*,noitemsep,topsep=0pt]
    \item \textbf{Inclusion Criteria:} Focused on post-contrast phases (e.g., arterial, venous, nephrographic), axial orientation, and diagnostic sequences (e.g., T1/T2, FLAIR, DWI for MRI).
    \item \textbf{Exclusion Criteria:} Series labeled as scout, localizer, pre-contrast, sagittal, coronal, or survey views.
    \item \textbf{Modality-Specific Rules:} Separate tailored keyword lists were used for MRI and CT scans to maximize phase relevance and minimize noise.
\end{itemize}
\medskip
After filtering, 716 high quality scans were retained from TCGA-KIRC and 191 from CPTAC-CCRCC. This resulted in 100 unique patients from TCGA and 35 from CPTAC, with an average of 7.2 and 5.5 usable scans per patient, respectively. Notably, MRI scans constituted the majority of the retained dataset (529 from TCGA, 137 from CPTAC), compared to CT scans (187 from TCGA, 54 from CPTAC).

\medskip 

\begin{table}[h!]
\centering
\caption{Summary statistics after CT/MRI scan filtering.}
\label{tab:ct_mri_summary_stats}
\begin{tabular}{lcc} 
\toprule 
\textbf{Metric} & \textbf{CPTAC-CCRCC} & \textbf{TCGA-KIRC} \\
\midrule 
Total original scans     & 814   & 2650 \\
Total filtered scans     & 191   & 716  \\
Percentage kept          & 23.5\% & 27.0\% \\
Unique patients          & 35    & 100  \\
Avg scans per patient    & 5.5   & 7.2  \\
CT scans                 & 54    & 187  \\
MRI scans                & 137   & 529  \\
\bottomrule 
\end{tabular}
\end{table}

\begin{figure}[h!]
\centering
\begin{subfigure}[t]{0.32\textwidth} 
    \centering
    \includegraphics[width=\textwidth]{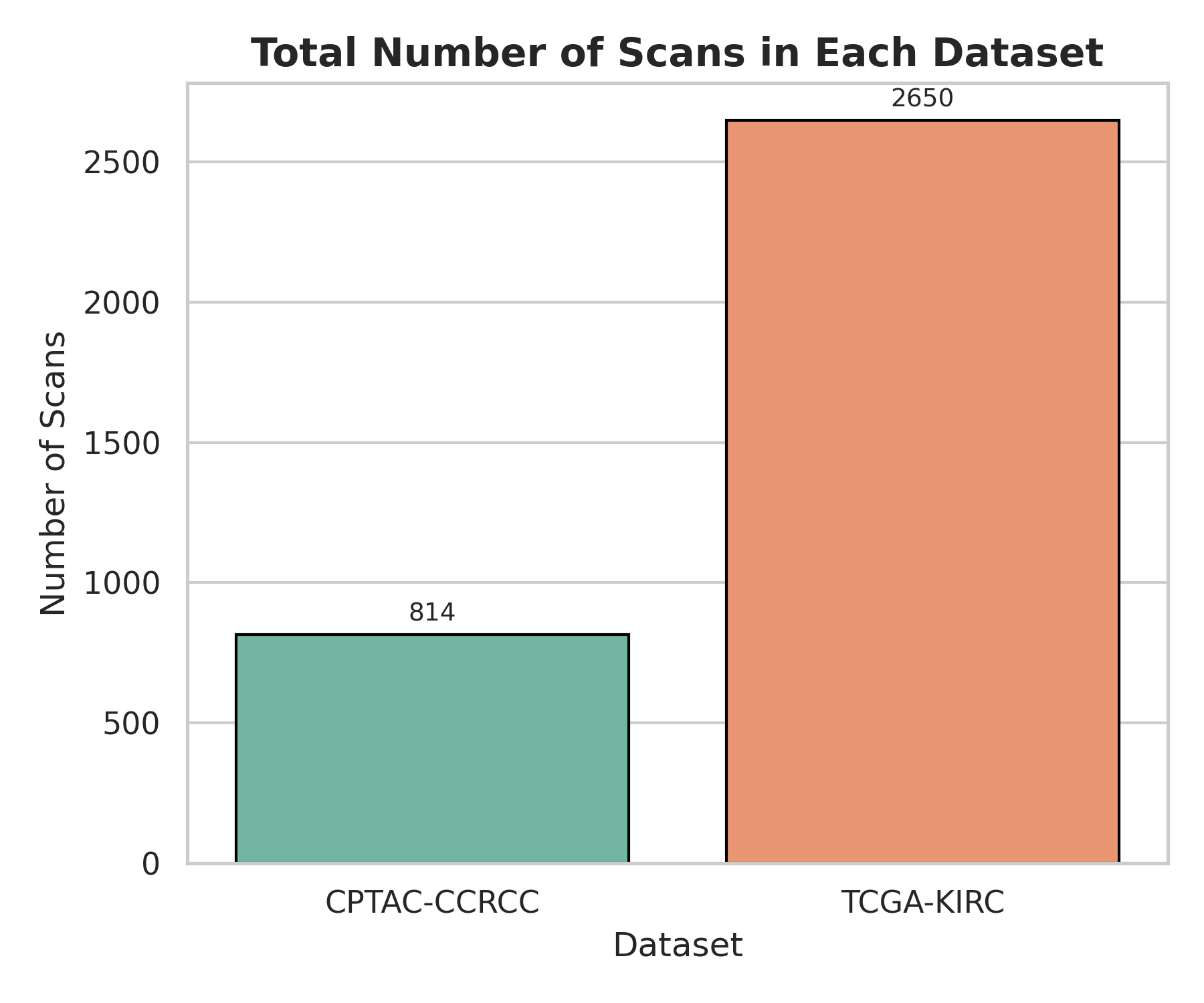}
    \caption{Total scans in each dataset}
    \label{fig:total_scans}
\end{subfigure}
\hfill 
\begin{subfigure}[t]{0.32\textwidth}
    \centering
    \includegraphics[width=\textwidth]{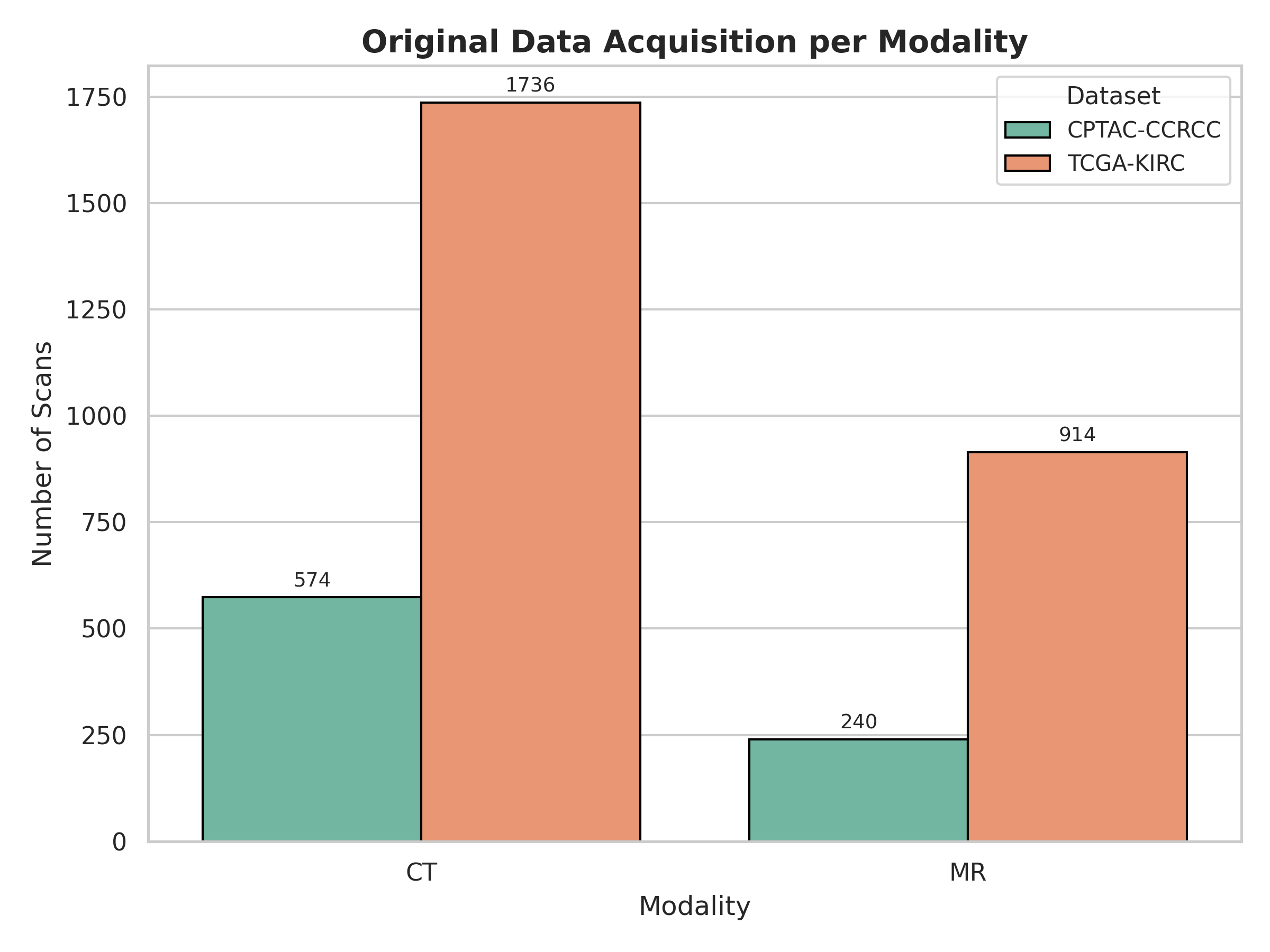}
    \caption{Original modality distribution}
    \label{fig:original_modality}
\end{subfigure}
\hfill
\begin{subfigure}[t]{0.32\textwidth}
    \centering
    \includegraphics[width=\textwidth]{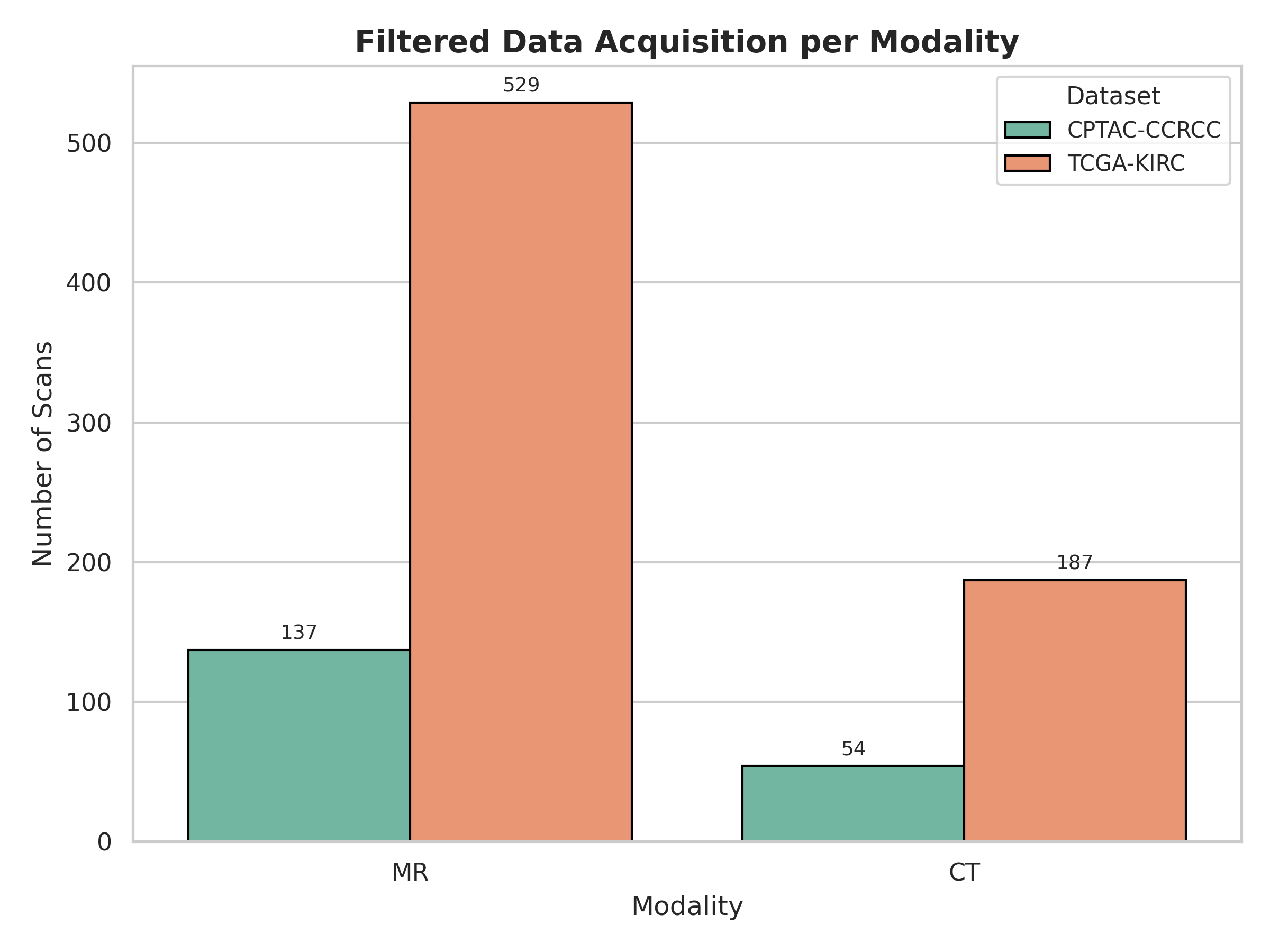}
    \caption{Filtered modality distribution}
    \label{fig:filtered_modality}
\end{subfigure}
\caption{Scan and modality distributions across CPTAC-CCRCC and TCGA-KIRC datasets before and after filtering.}
\label{fig:ct_mri_distributions} 
\end{figure}

\subsubsection{Preprocessing and Feature Extraction}
Each filtered DICOM series was reorganized into a structured directory grouped by \texttt{PatientID} and \texttt{SeriesInstanceUID}. Metadata mapping was created to associate each scan with its corresponding patient, ensuring alignment with labels and downstream fusion steps.
\medskip
For preprocessing and feature extraction:
\begin{itemize}[leftmargin=*,noitemsep,topsep=0pt]
    \item \textbf{Volume Normalization:} Each scan was clipped to the range $[-1024, 1024]$ Hounsfield Units (HU), then normalized using MeD3D statistics ($\mu = -158.58$, $\sigma = 324.70$).
    \item \textbf{Spatial Resampling:} Volumes were resized to a fixed shape of $56 \times 448 \times 448$ voxels to match the expected input of the pre-trained model.
    \item \textbf{Feature Extraction:} We utilized \texttt{resnet18} from MedicalNet (pre-trained on 23 medical datasets) as the backbone. The classification head was removed, and the final convolutional features were globally pooled to produce a single 512-dimensional embedding per scan.
    \item \textbf{Data Augmentation:} Gaussian noise was added to feature vectors during training to enhance generalization, particularly for underrepresented MRI samples.
\end{itemize}
Feature vectors were saved in \texttt{.npy} format per scan and grouped by patient folder for downstream aggregation.

\subsubsection{Patient-Level Embedding Aggregation}
For each patient, scan-level embeddings were aggregated via mean pooling to form a single 512-dimensional vector. This process generated the final input representation for the CT/MRI modality, saved as a CSV file containing 135 patient entries with consistent identifiers aligned across all modalities for multimodal fusion.

\subsubsection{Model Architecture and Training}
The network comprised two fully connected layers (512 $\rightarrow$ 256 $\rightarrow$ 128) followed by a final classification layer. A max-pooling operation across scans was employed to aggregate instance-level predictions into a patient-level output.

\medskip 
The training strategy included:
\begin{itemize}[leftmargin=*,noitemsep,topsep=0pt]
    \item \textbf{Loss Function:} Binary Cross-Entropy with \texttt{pos\_weight} to mitigate class imbalance.
    \item \textbf{Optimization:} Adam optimizer with weight decay and a fixed learning rate.
    \item \textbf{Batching:} \texttt{WeightedRandomSampler} was used to construct balanced mini-batches during training.
    \item \textbf{Epochs:} Models were trained for 70 epochs with early stopping based on validation loss.
\end{itemize}

\subsubsection{Inference and Representation}
During inference, per-scan outputs were aggregated using the maximum predicted survival probability to represent the patient's recurrence risk. Intermediate 128-dimensional embeddings were retained for downstream fusion and visualization.

\subsection{Whole-Slide Imaging (WSI)}
\label{WS}
This section details the acquisition, processing, and analysis of Whole-Slide Imaging (WSI) data for clear cell Renal Cell Carcinoma (ccRCC) patients, utilizing the optimized CLAM pipeline for weakly supervised learning.

\subsubsection{Data Acquisition and Preprocessing}
A total of 2,573 diagnostic and adjacent tissue WSI slides from 618 ccRCC patients were collected from The Cancer Genome Atlas (TCGA) and Clinical Proteomic Tumor Analysis Consortium (CPTAC) cohorts. Each case typically included two slides (tumor and adjacent tissue), with some cases having additional supplemental slides (3--12). All slides were formalin-fixed, paraffin-embedded (FFPE), stained with hematoxylin and eosin (H\&E), and digitized using clinical whole-slide scanners at either $20\times$ (0.5~$\mu$m/pixel) or $40\times$ (0.25~$\mu$m/pixel) resolution. Image formats included \texttt{.svs}, \texttt{.ndpi}, and \texttt{.tiff}.

\medskip
Preprocessing involved the following steps:
\medskip
\begin{itemize}[noitemsep,topsep=0pt]
    \item \textbf{Tissue Segmentation:} Each WSI was segmented using thresholding and contour filtering techniques, including area thresholds and hole filling. Slide-specific parameters were automatically logged.
    \item \textbf{Patch Extraction:} From the foreground tissue regions, $256 \times 256$ pixel patches were extracted at the highest available resolution.
    \item \textbf{Patch Storage:} Coordinates of the extracted patches were saved in \texttt{.h5} files, with optional tissue masks and stitched previews generated for visualization and quality assurance.
\end{itemize}

\subsubsection{Feature Extraction}
Patch-level features were extracted using the CLAM pipeline's \texttt{extract\_features\_fp.py} script with on-the-fly patch loading. We evaluated multiple pre-trained encoders:

\medskip
\begin{itemize}[noitemsep,topsep=0pt]
    \item \textbf{ResNet50:} The default encoder, producing 1024-dimensional embeddings.
    \item \textbf{UNI} and \textbf{CONCH:} State-of-the-art Vision Transformer (ViT)-based encoders from Mahmood Lab, yielding 1024-dimensional (UNI) and 512-dimensional (CONCH) representations, respectively.
\end{itemize}

\medskip
Extracted features were saved as \texttt{.pt} files for each slide, with each file containing a tensor of patch-level embeddings and associated metadata.

\subsubsection{Patient-Level Feature Aggregation}
To prepare WSI features for downstream modeling and multimodal fusion, patch-level information was aggregated into patient-level representations through a two-stage averaging process:
\medskip
\begin{enumerate}[noitemsep,topsep=0pt]
    \item For each \texttt{.pt} file corresponding to an individual WSI, patch-level embeddings were averaged to obtain a single slide-level feature vector.
    \item For patients with multiple WSIs, all slide-level vectors were further averaged to generate a single 1024-dimensional feature vector per patient.
\end{enumerate}
\medskip
The final patient-level feature matrix was saved as a CSV file named \texttt{wsi\_features.csv}, serving as the unified WSI modality input for subsequent experiments.


\subsubsection{Integration for Multimodal Fusion}
For downstream multimodal fusion, per-slide features were aggregated into a single 1024D or 512D vector (e.g., via attention-weighted mean). If a patient had multiple slides, the highest-attention slide was selected to represent the patient's WSI features.

\subsection{Multimodal Fusion Experiments}
\label{MF}
Our multimodal fusion pipeline was constructed in two stages:

\begin{enumerate}
    \item \textbf{Modality specific Baseline Models:}  
    Each modality EHR, CT/MRI, and WSI was modeled independently using a dedicated MLP classifier trained on pre-extracted embeddings. These baselines served as references for understanding the individual predictive power of each data stream. Performance metrics and qualitative visualizations (confusion matrices, precision-recall curves, UMAPs) were reported to assess classifier behavior.

    \item \textbf{Fusion Strategies:}  
    To exploit the complementary nature of multimodal features, we evaluated two integration strategies:
    \begin{itemize}
        \item \textit{Late Fusion:} Combining modality specific classifier outputs either through weighted averaging or a trainable fusion head.
        \item \textit{Early Fusion:} Concatenating or mean pooling projected embeddings into a shared representation for unified classification.
    \end{itemize}
\end{enumerate}

This staged approach allowed us to analyze each modality’s individual contribution before exploring synergistic effects via multimodal fusion. Quantitative comparisons are summarized in Tables~\ref{tab:fusion-results} and \ref{tab:overall-comparison}.



\subsubsection{Unimodal Baseline Classifiers}
Before applying fusion strategies, we trained separate baseline classifiers for each data modality to evaluate their individual predictive potential. For each modality EHR, CT/MRI, and WSI we extracted patient level features and trained an independent MLP classifier. The evaluation included confusion matrices, PR curves, and UMAP  \cite{mcinnes2018umap} visualizations to understand model behavior.

\bmhead{EHR Modality: Clinical and Genomic MLP Baseline}
Clinical and genomic features were preprocessed and passed through a 2-layer MLP classifier with dropout and label smoothing. The model was trained using a stratified split, and the best validation epoch was selected using balanced accuracy as the criterion.

\begin{table}[h!]
\centering
\caption{EHR MLP Baseline Performance}
\begin{tabular}{lcc}
\hline
\textbf{Metric} & \textbf{Score} \\
\hline
Balanced Accuracy   & 0.81 \\
F1 Score            & 0.77 \\
Precision           & 1.0 \\
Recall              & 0.63 \\
\hline
\end{tabular}
\label{tab:ehr_baseline}
\end{table}

\begin{figure}[h!]
    \centering
    \includegraphics[width=0.9\textwidth]{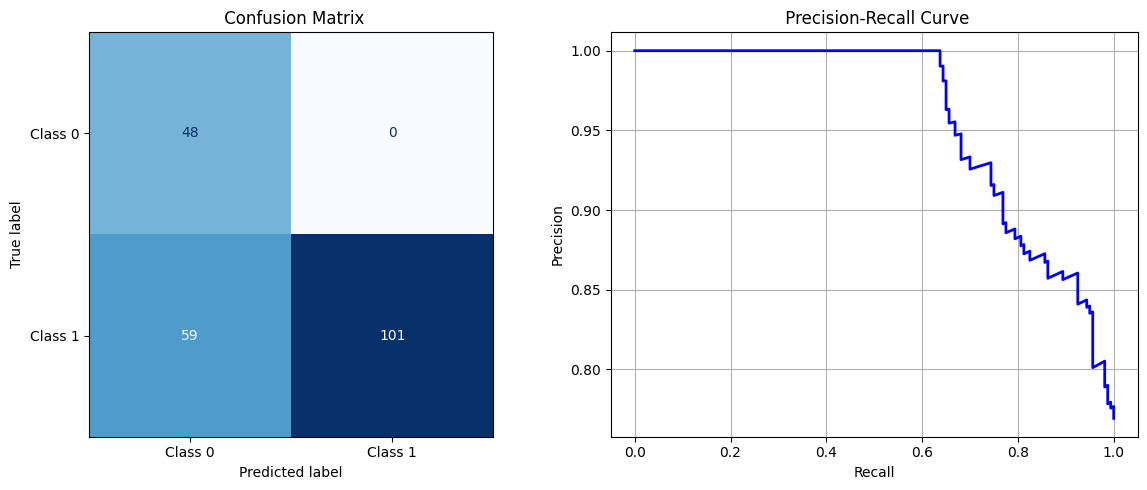}
    \caption{Left: Confusion matrix for the EHR classifier showing efficient class separation. Right: Precision-Recall curve with a smooth trend, validating probability calibration.}
    \label{fig:ehr_pr_curve}
\end{figure}

The confusion matrix and precision-recall curve in Figure~\ref{fig:ehr_pr_curve} offer deeper insight into model behavior. The classifier demonstrates strong precision and overall balanced prediction, though some recurrence cases were missed.

\begin{figure}[h!]
    \centering
    \includegraphics[width=0.75\textwidth]{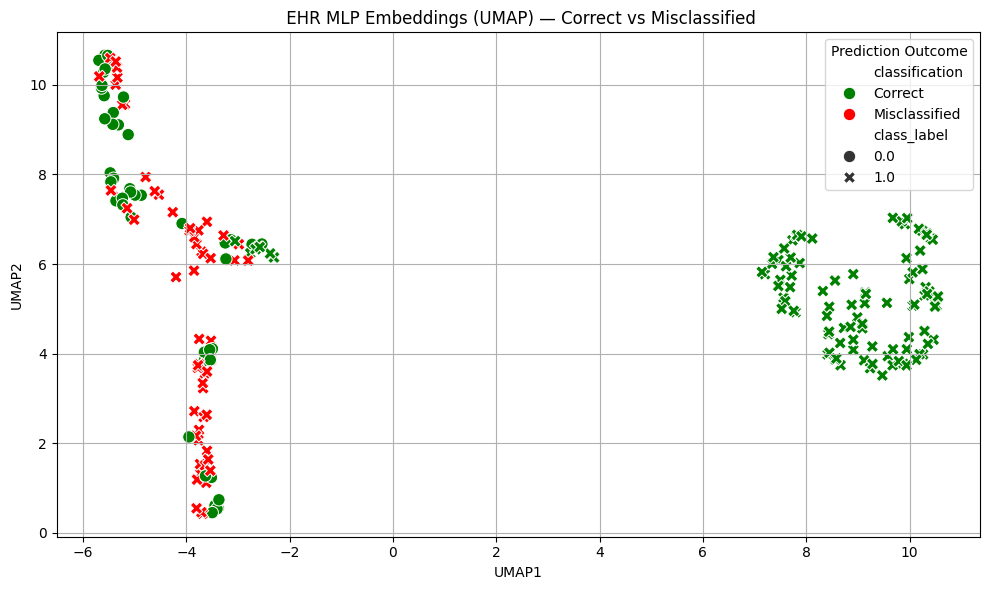}
    \caption{UMAP projection of the EHR model's intermediate layer. Green points denote correct predictions, and red points indicate misclassifications. Marker shape reflects ground-truth label.}
    \label{fig:ehr_umap}
\end{figure}

To better understand internal representations, UMAP was used to project high-dimensional fc1 layer outputs into 2D. As shown in Fig~\ref{fig:ehr_umap}, correctly classified patients form discernible clusters, while misclassified samples appear closer to class boundaries.

\begin{figure}[h!]
    \centering
    \includegraphics[width=0.6\textwidth]{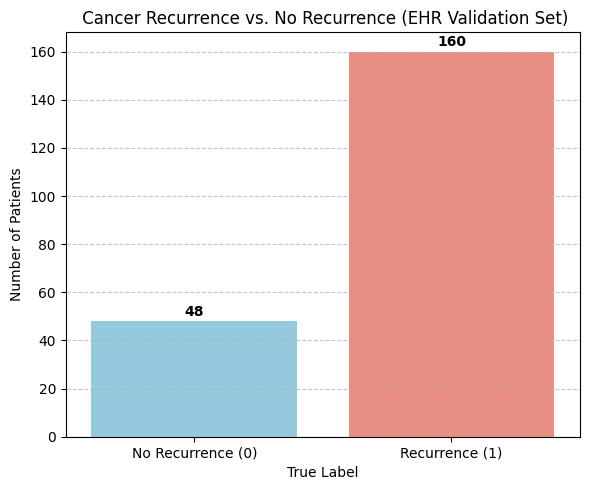}
    \caption{Cancer Recurrence vs. No Recurrence in the EHR validation set. A total of 208 patients were included, with 160 showing recurrence.}
    \label{fig:ehr_class_dist}
\end{figure}

Fig~\ref{fig:ehr_class_dist} presents the class distribution in the validation set. Among 208 patients, 160 had cancer recurrence (77\%), while 48 did not (23\%).

\bmhead{CT/MRI Modality: Radiological MLP Baseline}
Radiological features were extracted using a 3D-ResNet18 from the MedicalNet repository. After preprocessing and filtering diagnostic axial scans, These embeddings were fed into a lightweight MLP classifier for survival prediction.

\begin{table}[h!]
\centering
\caption{CT/MRI MLP Baseline Performance}
\begin{tabular}{lcc}
\hline
\textbf{Metric} & \textbf{Score} \\
\hline
Balanced Accuracy   & 0.86 \\
F1 Score            & 0.84 \\
Precision           & 1.00 \\
Recall              & 0.73 \\
\hline
\end{tabular}
\label{tab:ctmri_baseline}
\end{table}

\begin{figure}[h!]
    \centering
    \includegraphics[width=0.95\textwidth]{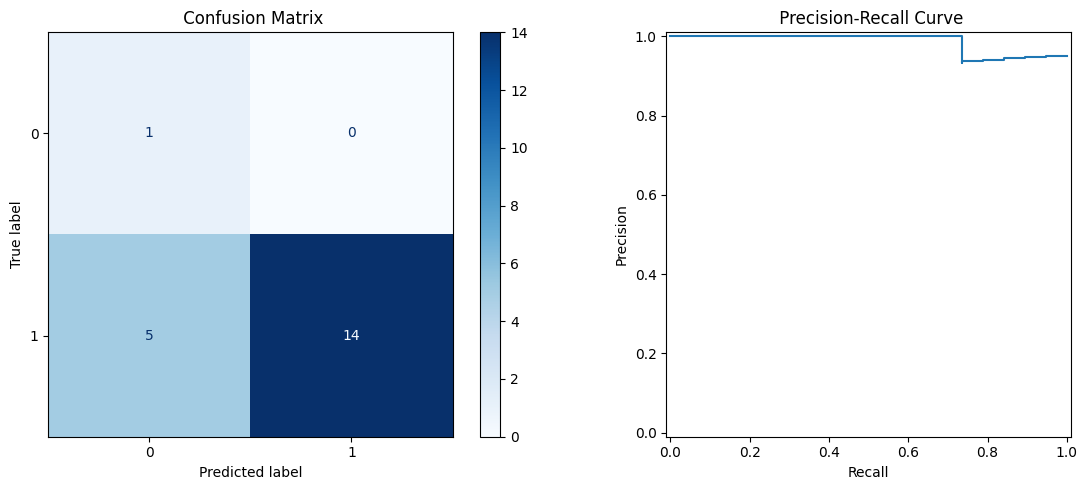}
    \caption{Confusion matrix (left) and precision-recall curve (right) for the CT/MRI MLP classifier. The classifier reliably detects recurrence cases with high precision.}
    \label{fig:ctmri_confusion_pr}
\end{figure}

Fig~\ref{fig:ctmri_confusion_pr} displays the confusion matrix and precision-recall curve for the CT/MRI classifier. The model achieved perfect precision, showing a strong ability to correctly identify recurrence, with a few false negatives.

\begin{figure}[h!]
    \centering
    \includegraphics[width=0.7\textwidth]{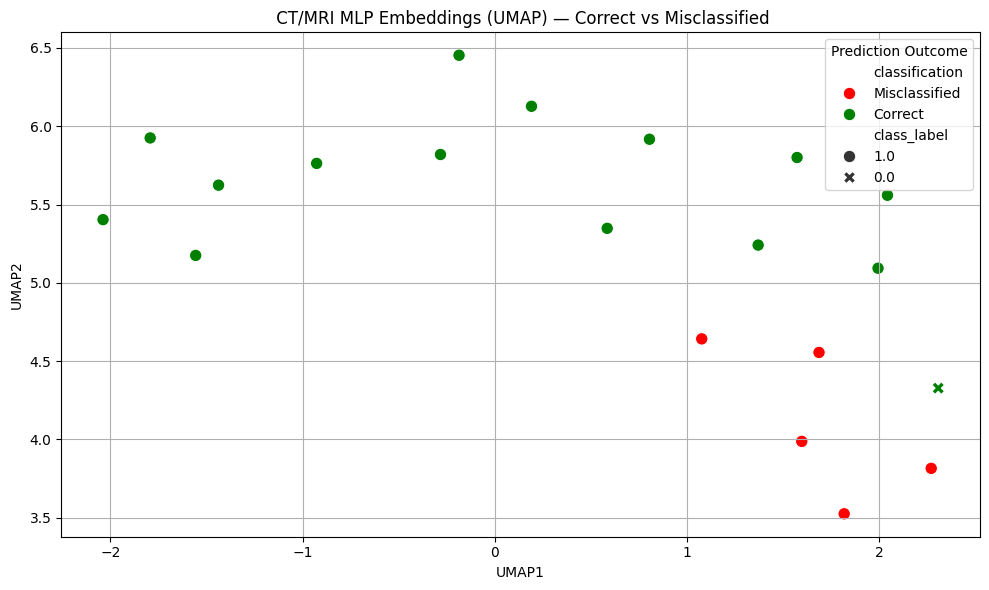}
    \caption{UMAP projection of CT/MRI intermediate features. Green points represent correctly classified patients, and red denote misclassified ones. Class label shapes help visualize decision boundary overlaps.}
    \label{fig:ctmri_umap}
\end{figure}

The UMAP projection in Figure~\ref{fig:ctmri_umap} illustrates how patient embeddings cluster in 2D space. Most recurrence cases form tight, separable clusters with clear decision boundaries, while a few misclassified samples fall near the margin.

\begin{figure}[h!]
    \centering
    \includegraphics[width=0.6\textwidth]{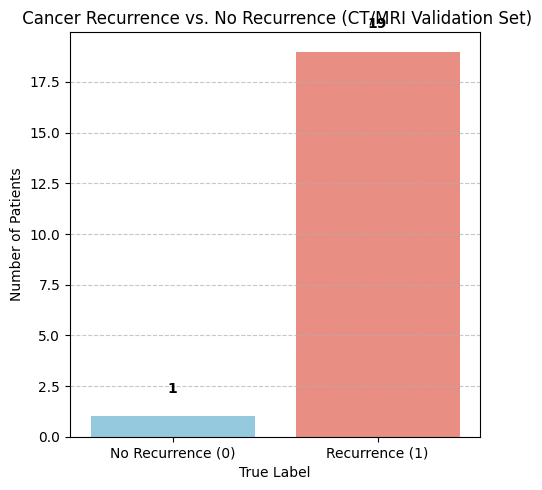}
    \caption{Cancer Recurrence vs. No Recurrence in the CT/MRI validation set. The dataset contains 20 patients, with a dominant recurrence class.}
    \label{fig:ctmri_class_dist}
\end{figure}

The distribution of recurrence labels is shown in Fig~\ref{fig:ctmri_class_dist}. Out of 20 patients in the validation set, 19 experienced cancer recurrence (95\%), and 1 had no recurrence (5\%). This indicates a strong skew in the validation cohort.

\bmhead{WSI Modality: Histopathology MLP Baseline}
Patch-level features were extracted from each WSI using the CLAM model with pretrained ResNet50 or CONCH encoders. Attention-based pooling was used to aggregate patch embeddings into a single 1024-dimensional vector per patient. These vectors were used to train a dedicated MLP classifier for recurrence prediction.

\begin{table}[h!]
\centering
\caption{WSI MLP Baseline Performance}
\begin{tabular}{lcc}
\toprule
\textbf{Metric} & \textbf{Score} \\
\midrule
Balanced Accuracy   & 0.70 \\
F1 Score            & 0.73 \\
Precision           & 0.95 \\
Recall              & 0.60 \\
\bottomrule
\end{tabular}
\label{tab:wsi_baseline}
\end{table}

\begin{figure}[h!]
    \centering
    \includegraphics[width=0.95\textwidth]{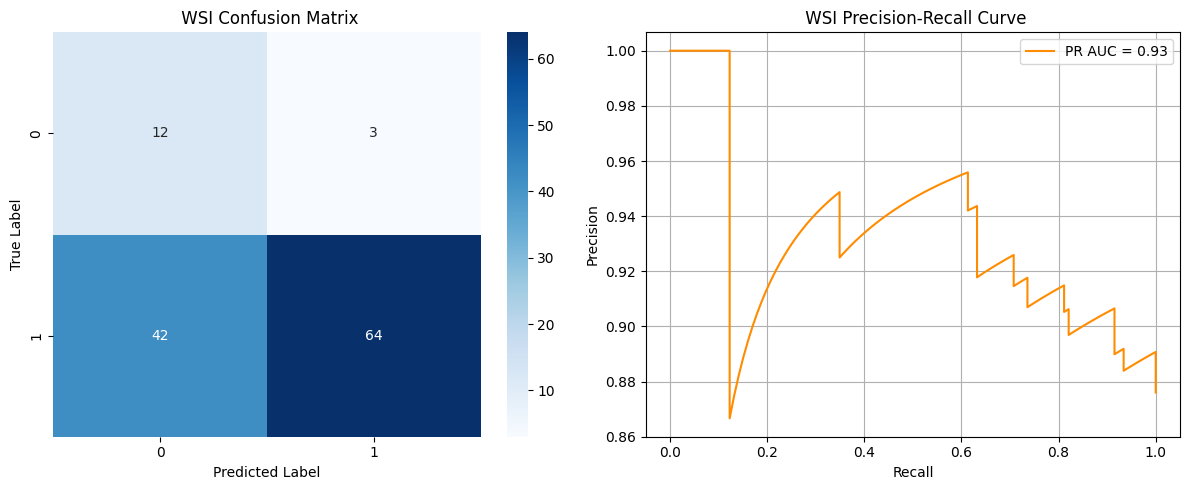}
    \caption{WSI modality: Confusion matrix (left) and precision-recall curve (right). The classifier demonstrates high confidence in recurrence predictions, with a PR curve of 0.95.}
    \label{fig:wsi_confusion_pr}
\end{figure}

Fig~\ref{fig:wsi_confusion_pr} shows that the model exhibits strong discriminative power in recurrence classification. It achieves a high F1-score with balanced precision and recall, evident from the PR curve and confusion matrix.

\begin{figure}[h!]
    \centering
    \includegraphics[width=0.85\textwidth]{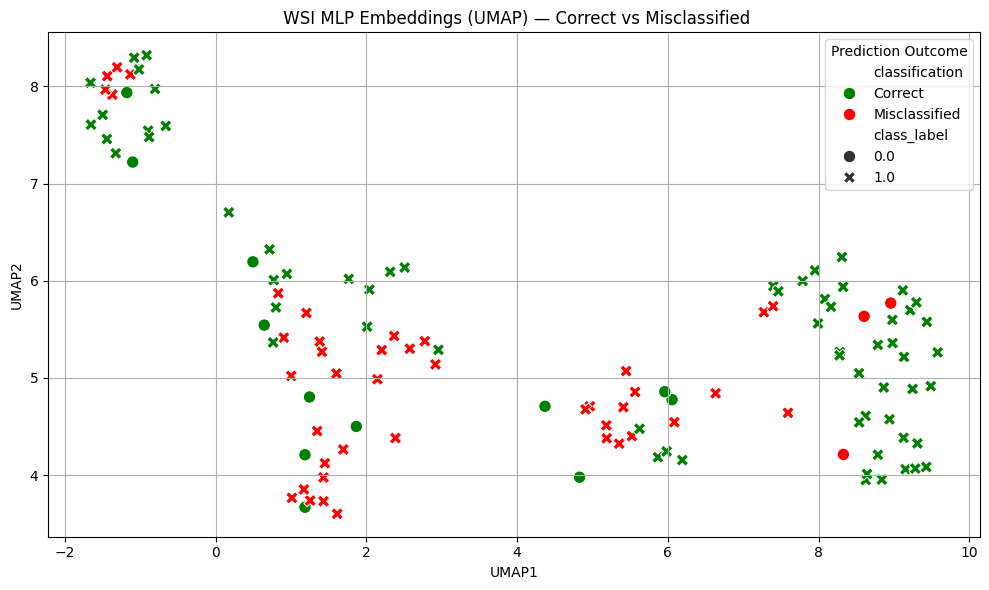}
    \caption{WSI modality: UMAP projection of hidden layer embeddings colored by prediction correctness. Green points denote correctly predicted samples, and red indicate errors. Shape corresponds to ground-truth class.}
    \label{fig:wsi_umap}
\end{figure}

Fig~\ref{fig:wsi_umap} visualizes UMAP-reduced embeddings of histopathology samples. Most correctly classified samples are well separated in latent space, while misclassified ones cluster near boundaries highlighting potential ambiguity in certain slides.

\begin{figure}[h!]
    \centering
    \includegraphics[width=0.6\textwidth]{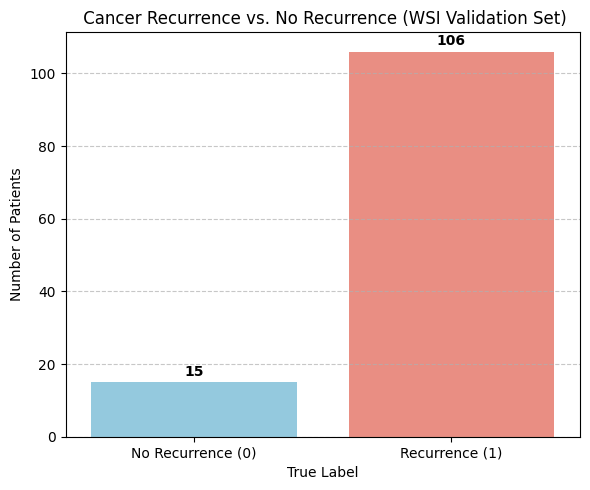}
    \caption{Cancer Recurrence vs. No Recurrence (WSI Validation Set). The cohort is predominantly composed of recurrence cases.}
    \label{fig:wsi_class_dist}
\end{figure}

To provide context on class proportions, Fig~\ref{fig:wsi_class_dist} shows the recurrence label distribution in the validation set. Out of 121 patients, 106 (87.6\%) had recurrence, while 15 (12.4\%) did not.

\subsubsection{Fusion Strategies}

To exploit the complementary nature of multimodal data, we evaluated both \textit{early fusion} and \textit{late fusion} techniques:

\begin{itemize}
    \item \textbf{Late Fusion (Weighted Sum)}: Probabilities from modality specific models were combined using weights proportional to their balanced accuracies.
    \item \textbf{Early Fusion (Concatenation)}: Feature embeddings from each modality were concatenated and passed to a unified classifier.
\end{itemize}

\subsubsection{Fusion Performance outcome}
Table \ref{tab:fusion-results} presents the performance results of two multimodal fusion strategies, Late Fusion (Weighted Sum) and Early Fusion (concatenation), in four evaluation metrics: Balanced Accuracy, F1 Score, Precision, and Recall. The Late Fusion strategy achieved a balanced accuracy of 0.667, an F1 score of 0.800, a perfect precision of 1.000, and a recall of 0.667. In comparison, the Early Fusion strategy outperformed the Late Fusion approach, with a balanced accuracy of 0.833, an F1 score of 0.983, a precision of 0.967, and a perfect recall of 1.000, highlighting its overall superior performance across metrics.
\begin{table}[h!]
\centering
\caption{Performance of Multimodal Fusion Strategies}
\begin{tabular}{lcccc}
\hline
\textbf{Model} & \textbf{Balanced Accuracy} & \textbf{F1 Score} & \textbf{Precision} & \textbf{Recall} \\
\hline
\textbf{Late Fusion (Weighted Sum)}    & 0.667 & 0.800 & \textbf{1.000} & 0.667 \\
\textbf{Early Fusion (Concatenation)}  & \textbf{0.833} & \textbf{0.983} & 0.967 & \textbf{1.000} \\
\hline
\end{tabular}
\label{tab:fusion-results}
\end{table}
\subsubsection{Overall Comparison of All Experiments}
For clarity and completeness, Table~\ref{tab:overall-comparison} presents a unified summary of all experimental results across the baseline, and fusion models. It highlights the benefit of integrating multimodal data.

Table 11 presents a performance comparison of various baseline and fusion models based on four evaluation metrics: Balanced Accuracy, F1 Score, Precision, and Recall. The baseline models include EHR MLP, CT/MRI MLP, and WSI MLP, with the CT/MRI MLP baseline achieving the highest Balanced Accuracy (0.868) and F1 Score (0.848). The fusion models, which include Late Fusion (Weighted Sum) and Early Fusion (Concatenation), offer a mix of performance. The Late Fusion model achieves a Balanced Accuracy of 0.667 and F1 Score of 0.800, while the Early Fusion model outperforms all others with the highest Balanced Accuracy (0.833), F1 Score (0.983), and Recall (1.000), demonstrating superior performance in terms of precision and recall across the models.

\begin{table}[h!]
\centering
\caption{Unified Performance Comparison: Baselines, and Fusion Models}
\begin{tabular}{lcccc}
\hline
\textbf{Model} & \textbf{Balanced Accuracy} & \textbf{F1 Score} & \textbf{Precision} & \textbf{Recall} \\
\hline
\textbf{EHR MLP Baseline}              & 0.816 & 0.774 & \textbf{1.000} & 0.631 \\
\textbf{CT/MRI MLP Baseline}           & 0.868 & 0.848 & \textbf{1.000} & 0.737 \\
\textbf{WSI MLP Baseline}              & 0.702 & 0.740 & 0.955 & 0.604 \\
\hline
\textbf{Late Fusion (Weighted Sum)}    & 0.667 & 0.800 & \textbf{1.000} & 0.667 \\
\textbf{Early Fusion (Concatenation)}  & \textbf{0.833} & \textbf{0.983} & 0.967 & \textbf{1.000} \\
\hline
\end{tabular}
\label{tab:overall-comparison}
\end{table}

\begin{quote}
\textit{Note: Bolded values in the tables represent the best performance achieved for each corresponding metric.}
\end{quote}

\section{Discussion}\label{sec:discussion}
The conducted experiments clearly underline the significant advantages of multimodal data fusion in enhancing performance. Each modality independently contributed meaningful signals, with notable results from CT/MRI achieving a balanced accuracy of 86.8\%, and EHR data showing perfect precision. However, fusion strategies, particularly the early fusion approach, yielded superior performance metrics when considering the integration of these modalities.

The early fusion model demonstrated the highest performance, yielding a balanced accuracy of 83.3\%, an F1 score of 0.983, and a perfect recall rate of 100\%. The mathematical formulation of balanced accuracy ($Acc_{\text{bal}}$) is given by:

\[
Acc_{\text{bal}} = \frac{1}{2} \left( \frac{TP}{TP + FN} + \frac{TN}{TN + FP} \right)
\]

where $TP$, $TN$, $FP$, and $FN$ denote true positives, true negatives, false positives, and false negatives, respectively. These results underscore the model's robustness in correctly identifying recurrence cases, where recall ($R$) is defined as:

\[
R = \frac{TP}{TP + FN} = 1 \quad \text{(Perfect recall)}
\]

On the other hand, the late fusion strategy achieved perfect precision ($P = 1$) but displayed a reduced balanced accuracy of 66.7\%. This suggests a potential trade-off between sensitivity (recall) and specificity (precision) in certain fusion schemes, which can be further explored using the following precision-recall relationship:

\[
P = \frac{TP}{TP + FP} = 1 \quad \text{(Perfect precision)}
\]

This discrepancy indicates that while the late fusion model excels in correctly identifying positive instances, its ability to discriminate between negative and positive cases, particularly in terms of balanced accuracy, is less optimal.

\bmhead{Outcome:} These findings support the hypothesis that multimodal integration—leveraging radiological, pathological, and clinical data—substantially improves prognostic performance. The early fusion model, in particular, highlights the importance of incorporating complementary information in clinical settings. These results justify the need for further clinical validation studies to confirm the robustness of these fusion strategies in real-world scenarios.

\section{Conclusion}\label{sec:Con}
This study introduces a comprehensive and interpretable deep learning pipeline designed for recurrence prediction in ccRCC using multimodal data. By independently modeling the features from EHR, CT/MRI, and WSI, we demonstrated the individual prognostic value of each modality in predicting patient outcomes. Notably, fusion through early concatenation substantially enhanced the overall performance, achieving an impressive 98.3\% F1 Score with perfect recall. These results underscore the potential of integrating multimodal data for personalized oncology approaches. Moreover, this work provides a solid foundation for further exploration of data-driven precision medicine, paving the way for innovative methodologies in the clinical prediction of cancer recurrence. Future advancements in this area could lead to more accurate and individualized treatment strategies, improving patient outcomes in oncology.
\bmhead{Future Work}
To further enhance the clinical applicability and scientific rigor of multimodal recurrence prediction, the study recommend the following directions:
\begin{itemize}\label{sec:future_work}
    \item \textbf{Advanced fusion architectures:} Implement attention based or transformer fusion models to dynamically learn modality relevance and interactions.
    \item \textbf{Handling missing modalities:} Develop models capable of inferring from partial inputs using uncertainty aware fusion or modality dropout techniques.
    \item \textbf{Explainability and trust:} Integrate SHAP, Grad-CAM, or attention heatmaps to provide transparency and aid clinical interpretation.
    \item \textbf{External and prospective validation:} Test the pipeline on multi institutional datasets to assess generalization and readiness for real-world deployment.
    \item \textbf{Joint representation learning:} Move toward joint multimodal embedding spaces through contrastive, self-supervised, or variational learning methods.
\end{itemize}
\backmatter

\section*{Declarations}\label{sec:declar}
\bmhead{Ethics approval and consent to participate}
Approved and Not applicable
\bmhead{Consent for publication}
Not applicable
\bmhead{Data availability}
Data will be made available on request.
\bmhead{Funding}
No funding
\bmhead{Declaration of competing interest}
The authors declare that they have no known competing financial interests or personal relationships that could have appeared to influence the work reported in this paper.
\bmhead{CRediT authorship contribution statement}
Hasaan Maqsood \& Saif Ur Rehman Khan: Conceptualization, Data curation, Methodology, Software, Validation, Writing original draft, and  review \& editing. 

\section*{Abbreviations and Definitions}\label{sec:abbrev}

\begin{table}[htbp]
\centering
\caption{Comprehensive Abbreviations and Definitions}
\label{tab:all_abbreviations}
\begin{tabular}{lp{0.75\linewidth}}
\toprule
\textbf{Abbreviation} & \textbf{Definition} \\
\midrule
\textbf{Clinical Terms} \\
ccRCC & Clear Cell Renal Cell Carcinoma \\
AJCC & American Joint Committee on Cancer staging system \\
pT/pN/pM & Pathological Tumor/Node/Metastasis stage \\
VHL & Von Hippel-Lindau tumor suppressor gene \\
PBRM1 & Polybromo-1 (chromatin remodeling gene) \\
TTN & Titin (structural protein gene) \\
H\&E & Hematoxylin and Eosin (histopathology stain) \\

\textbf{Imaging Modalities} \\
WSI & Whole Slide Image (digital pathology) \\
CT & Computed Tomography \\
MRI & Magnetic Resonance Imaging \\
DICOM & Digital Imaging and Communications in Medicine \\
HU & Hounsfield Units (CT intensity measurement) \\
FFPE & Formalin-Fixed Paraffin-Embedded \\

\textbf{Methods \& Models} \\
MIL & Multiple Instance Learning \\
CLAM & Clustering-constrained Attention MIL \\
MeD3D & Medical 3D Deep Learning framework \\
SMOTE & Synthetic Minority Over-sampling Technique \\
MLP & Multilayer Perceptron \\
CNN & Convolutional Neural Network \\
ViT & Vision Transformer \\

\textbf{Datasets} \\
TCGA-KIRC & The Cancer Genome Atlas Kidney Renal Clear Cell Carcinoma \\
CPTAC-CCRCC & Clinical Proteomic Tumor Analysis Consortium Clear Cell RCC \\
TCIA & The Cancer Imaging Archive \\

\textbf{Codings} \\
Gender & 1=Male, 0=Female \\
Vital Status & 1=Deceased, 0=Living \\
Mutation & 1=Present, 0=Absent, -1=Unknown \\
\bottomrule
\end{tabular}

\smallskip
\footnotesize
\textit{Note:} Comprehensive abbreviations used throughout the MeD-3D multimodal framework.
\end{table}

\newpage

\bibliography{sn-bibliography}


\begin{thebibliography}{25}
\ifx \bisbn   \undefined \def \bisbn  #1{ISBN #1}\fi
\ifx \binits  \undefined \def \binits#1{#1}\fi
\ifx \bauthor  \undefined \def \bauthor#1{#1}\fi
\ifx \batitle  \undefined \def \batitle#1{#1}\fi
\ifx \bjtitle  \undefined \def \bjtitle#1{#1}\fi
\ifx \bvolume  \undefined \def \bvolume#1{\textbf{#1}}\fi
\ifx \byear  \undefined \def \byear#1{#1}\fi
\ifx \bissue  \undefined \def \bissue#1{#1}\fi
\ifx \bfpage  \undefined \def \bfpage#1{#1}\fi
\ifx \blpage  \undefined \def \blpage #1{#1}\fi
\ifx \burl  \undefined \def \burl#1{\textsf{#1}}\fi
\ifx \doiurl  \undefined \def \doiurl#1{\url{https://doi.org/#1}}\fi
\ifx \betal  \undefined \def \betal{\textit{et al.}}\fi
\ifx \binstitute  \undefined \def \binstitute#1{#1}\fi
\ifx \binstitutionaled  \undefined \def \binstitutionaled#1{#1}\fi
\ifx \bctitle  \undefined \def \bctitle#1{#1}\fi
\ifx \beditor  \undefined \def \beditor#1{#1}\fi
\ifx \bpublisher  \undefined \def \bpublisher#1{#1}\fi
\ifx \bbtitle  \undefined \def \bbtitle#1{#1}\fi
\ifx \bedition  \undefined \def \bedition#1{#1}\fi
\ifx \bseriesno  \undefined \def \bseriesno#1{#1}\fi
\ifx \blocation  \undefined \def \blocation#1{#1}\fi
\ifx \bsertitle  \undefined \def \bsertitle#1{#1}\fi
\ifx \bsnm \undefined \def \bsnm#1{#1}\fi
\ifx \bsuffix \undefined \def \bsuffix#1{#1}\fi
\ifx \bparticle \undefined \def \bparticle#1{#1}\fi
\ifx \barticle \undefined \def \barticle#1{#1}\fi
\bibcommenthead
\ifx \bconfdate \undefined \def \bconfdate #1{#1}\fi
\ifx \botherref \undefined \def \botherref #1{#1}\fi
\ifx \url \undefined \def \url#1{\textsf{#1}}\fi
\ifx \bchapter \undefined \def \bchapter#1{#1}\fi
\ifx \bbook \undefined \def \bbook#1{#1}\fi
\ifx \bcomment \undefined \def \bcomment#1{#1}\fi
\ifx \oauthor \undefined \def \oauthor#1{#1}\fi
\ifx \citeauthoryear \undefined \def \citeauthoryear#1{#1}\fi
\ifx \endbibitem  \undefined \def \endbibitem {}\fi
\ifx \bconflocation  \undefined \def \bconflocation#1{#1}\fi
\ifx \arxivurl  \undefined \def \arxivurl#1{\textsf{#1}}\fi
\csname PreBibitemsHook\endcsname

\bibitem[\protect\citeauthoryear{{World Health Organization}}{2020}]{who2020}
\begin{botherref}
\oauthor{\bsnm{{World Health Organization}}}:
Cancer Fact Sheet
(2020).
\url{https://www.who.int/news-room/fact-sheets/detail/cancer}
\end{botherref}
\endbibitem

\bibitem[\protect\citeauthoryear{Waqas et~al.}{2024}]{waqas2024multimodal}
\begin{barticle}
\bauthor{\bsnm{Waqas}, \binits{A.}},
\bauthor{\bsnm{Tripathi}, \binits{A.}},
\bauthor{\bsnm{Ramachandran}, \binits{R.P.}},
\bauthor{\bsnm{Stewart}, \binits{P.A.}},
\bauthor{\bsnm{Rasool}, \binits{G.}}:
\batitle{Multimodal data integration for oncology in the era of deep neural networks: a review}.
\bjtitle{Frontiers in Artificial Intelligence}
\bvolume{7},
\bfpage{1408843}
(\byear{2024})
\doiurl{10.3389/frai.2024.1408843}
\end{barticle}
\endbibitem

\bibitem[\protect\citeauthoryear{Linehan and Ricketts}{2019}]{linehan2019cancer}
\begin{barticle}
\bauthor{\bsnm{Linehan}, \binits{W.M.}},
\bauthor{\bsnm{Ricketts}, \binits{C.J.}}:
\batitle{The cancer genome atlas of renal cell carcinoma: findings and clinical implications}.
\bjtitle{Nature Reviews Urology}
\bvolume{16}(\bissue{9}),
\bfpage{539}--\blpage{552}
(\byear{2019})
\end{barticle}
\endbibitem

\bibitem[\protect\citeauthoryear{Kase et~al.}{2023}]{kase2023clear}
\begin{barticle}
\bauthor{\bsnm{Kase}, \binits{A.M.}},
\bauthor{\bsnm{George}, \binits{D.J.}},
\bauthor{\bsnm{Ramalingam}, \binits{S.}}:
\batitle{Clear cell renal cell carcinoma: from biology to treatment}.
\bjtitle{Cancers}
\bvolume{15}(\bissue{3}),
\bfpage{665}
(\byear{2023})
\doiurl{10.3390/cancers15030665}
\end{barticle}
\endbibitem

\bibitem[\protect\citeauthoryear{Chong et~al.}{2025}]{chong2025establishing}
\begin{barticle}
\bauthor{\bsnm{Chong}, \binits{Y.}},
\bauthor{\bsnm{Zhou}, \binits{H.}},
\bauthor{\bsnm{Zhang}, \binits{P.}},
\bauthor{\bsnm{Xue}, \binits{L.}},
\bauthor{\bsnm{Du}, \binits{Q.}},
\bauthor{\bsnm{Chong}, \binits{T.}},
\bauthor{\bsnm{Wang}, \binits{Z.}}:
\batitle{Establishing cm0 (i+) stage criteria in localized renal cell carcinoma based on postoperative circulating tumor cells monitoring}.
\bjtitle{BMC cancer}
\bvolume{25}(\bissue{1}),
\bfpage{436}
(\byear{2025})
\end{barticle}
\endbibitem

\bibitem[\protect\citeauthoryear{Khan}{2025}]{khan2025multi}
\begin{barticle}
\bauthor{\bsnm{Khan}, \binits{S.U.R.}}:
\batitle{Multi-level feature fusion network for kidney disease detection}.
\bjtitle{Computers in Biology and Medicine}
\bvolume{191},
\bfpage{110214}
(\byear{2025})
\end{barticle}
\endbibitem

\bibitem[\protect\citeauthoryear{Khan and Khan}{2025}]{khan2025detection}
\begin{botherref}
\oauthor{\bsnm{Khan}, \binits{S.U.R.}},
\oauthor{\bsnm{Khan}, \binits{Z.}}:
Detection of abnormal cardiac rhythms using feature fusion technique with heart sound spectrograms.
Journal of Bionic Engineering,
1--20
(2025)
\end{botherref}
\endbibitem

\bibitem[\protect\citeauthoryear{Khan et~al.}{2025}]{khan2025robust}
\begin{botherref}
\oauthor{\bsnm{Khan}, \binits{S.U.R.}},
\oauthor{\bsnm{Asim}, \binits{M.N.}},
\oauthor{\bsnm{Vollmer}, \binits{S.}},
\oauthor{\bsnm{Dengel}, \binits{A.}}:
Robust \& precise knowledge distillation-based novel context-aware predictor for disease detection in brain and gastrointestinal.
arXiv preprint arXiv:2505.06381
(2025)
\end{botherref}
\endbibitem

\bibitem[\protect\citeauthoryear{Khan et~al.}{2024}]{khan2024glnet}
\begin{barticle}
\bauthor{\bsnm{Khan}, \binits{S.U.R.}},
\bauthor{\bsnm{Zhao}, \binits{M.}},
\bauthor{\bsnm{Asif}, \binits{S.}},
\bauthor{\bsnm{Chen}, \binits{X.}},
\bauthor{\bsnm{Zhu}, \binits{Y.}}:
\batitle{Glnet: global--local cnn's-based informed model for detection of breast cancer categories from histopathological slides}.
\bjtitle{The Journal of Supercomputing}
\bvolume{80}(\bissue{6}),
\bfpage{7316}--\blpage{7348}
(\byear{2024})
\end{barticle}
\endbibitem

\bibitem[\protect\citeauthoryear{Iseke et~al.}{2023}]{Iseke2023}
\begin{barticle}
\bauthor{\bsnm{Iseke}, \binits{S.}},
\bauthor{\bsnm{Zeevi}, \binits{T.}},
\bauthor{\bsnm{Kucukkaya}, \binits{A.S.}},
\bauthor{\bsnm{Raju}, \binits{R.}},
\bauthor{\bsnm{Gross}, \binits{M.}},
\bauthor{\bsnm{Haider}, \binits{S.P.}},
\bauthor{\bsnm{Petukhova-Greenstein}, \binits{A.}},
\bauthor{\bsnm{Kuhn}, \binits{T.N.}},
\bauthor{\bsnm{Lin}, \binits{M.}},
\bauthor{\bsnm{Nowak}, \binits{M.}},
\bauthor{\bsnm{Cooper}, \binits{K.}}:
\batitle{Machine learning models for prediction of posttreatment recurrence in early-stage hepatocellular carcinoma using pretreatment clinical and mri features: a proof-of-concept study}.
\bjtitle{American Journal of Roentgenology}
\bvolume{220}(\bissue{2}),
\bfpage{245}--\blpage{255}
(\byear{2023})
\end{barticle}
\endbibitem

\bibitem[\protect\citeauthoryear{Wang et~al.}{2023}]{Wang2023}
\begin{barticle}
\bauthor{\bsnm{Wang}, \binits{H.}},
\bauthor{\bsnm{Zhang}, \binits{M.}},
\bauthor{\bsnm{Miao}, \binits{J.}},
\bauthor{\bsnm{Hou}, \binits{F.}},
\bauthor{\bsnm{Chen}, \binits{Y.}},
\bauthor{\bsnm{Huang}, \binits{Y.}},
\bauthor{\bsnm{Yang}, \binits{L.}},
\bauthor{\bsnm{Yang}, \binits{S.}},
\bauthor{\bsnm{Huang}, \binits{C.}},
\bauthor{\bsnm{Song}, \binits{Y.}},
\bauthor{\bsnm{Niu}, \binits{H.}}:
\batitle{Deep learning signature based on multiphase enhanced ct for bladder cancer recurrence prediction: a multi-center study}.
\bjtitle{EClinicalMedicine}
\bvolume{66},
\bfpage{101799}
(\byear{2023})
\end{barticle}
\endbibitem

\bibitem[\protect\citeauthoryear{Gu et~al.}{2023}]{Gu2023}
\begin{barticle}
\bauthor{\bsnm{Gu}, \binits{W.J.}},
\bauthor{\bsnm{Liu}, \binits{Z.}},
\bauthor{\bsnm{Yang}, \binits{Y.}},
\bauthor{\bsnm{Zhang}, \binits{X.}},
\bauthor{\bsnm{Chen}, \binits{L.}},
\bauthor{\bsnm{Wan}, \binits{F.}},
\bauthor{\bsnm{Liu}, \binits{X.H.}},
\bauthor{\bsnm{Chen}, \binits{Z.}},
\bauthor{\bsnm{Kong}, \binits{Y.}},
\bauthor{\bsnm{Dai}, \binits{B.}}:
\batitle{A deep learning model, nafnet, predicts adverse pathology and recurrence in prostate cancer using mris}.
\bjtitle{NPJ Precision Oncology}
\bvolume{7}(\bissue{1}),
\bfpage{134}
(\byear{2023})
\end{barticle}
\endbibitem

\bibitem[\protect\citeauthoryear{Cepeda et~al.}{2023}]{Cepeda2023}
\begin{barticle}
\bauthor{\bsnm{Cepeda}, \binits{S.}},
\bauthor{\bsnm{Luppino}, \binits{L.T.}},
\bauthor{\bsnm{P{\'e}rez-N{\'u}{\~n}ez}, \binits{A.}},
\bauthor{\bsnm{Solheim}, \binits{O.}},
\bauthor{\bsnm{Garc{\'i}a-Garc{\'i}a}, \binits{S.}},
\bauthor{\bsnm{Velasco-Casares}, \binits{M.}},
\bauthor{\bsnm{Karlberg}, \binits{A.}},
\bauthor{\bsnm{Eikenes}, \binits{L.}},
\bauthor{\bsnm{Sarabia}, \binits{R.}},
\bauthor{\bsnm{Arrese}, \binits{I.}},
\bauthor{\bsnm{Zamora}, \binits{T.}}:
\batitle{Predicting regions of local recurrence in glioblastomas using voxel-based radiomic features of multiparametric postoperative mri}.
\bjtitle{Cancers}
\bvolume{15}(\bissue{6}),
\bfpage{1894}
(\byear{2023})
\end{barticle}
\endbibitem

\bibitem[\protect\citeauthoryear{Subramanian et~al.}{2020}]{subramanian2020multimodal}
\begin{bchapter}
\bauthor{\bsnm{Subramanian}, \binits{V.}},
\bauthor{\bsnm{Do}, \binits{M.N.}},
\bauthor{\bsnm{Syeda-Mahmood}, \binits{T.}}:
\bctitle{Multimodal fusion of imaging and genomics for lung cancer recurrence prediction}.
In: \bbtitle{2020 IEEE 17th International Symposium on Biomedical Imaging (ISBI)},
pp. \bfpage{804}--\blpage{808}
(\byear{2020}).
\bcomment{IEEE}
\end{bchapter}
\endbibitem

\bibitem[\protect\citeauthoryear{Ren et~al.}{2023}]{Ren2023}
\begin{barticle}
\bauthor{\bsnm{Ren}, \binits{J.}},
\bauthor{\bsnm{Zhai}, \binits{X.}},
\bauthor{\bsnm{Yin}, \binits{H.}},
\bauthor{\bsnm{Zhou}, \binits{F.}},
\bauthor{\bsnm{Hu}, \binits{Y.}},
\bauthor{\bsnm{Wang}, \binits{K.}},
\bauthor{\bsnm{Yan}, \binits{R.}},
\bauthor{\bsnm{Han}, \binits{D.}}:
\batitle{Multimodality mri radiomics based on machine learning for identifying true tumor recurrence and treatment-related effects in patients with postoperative glioma}.
\bjtitle{Neurology and Therapy}
\bvolume{12}(\bissue{5}),
\bfpage{1729}--\blpage{1743}
(\byear{2023})
\end{barticle}
\endbibitem

\bibitem[\protect\citeauthoryear{Qiu et~al.}{2022}]{Qiu2022}
\begin{barticle}
\bauthor{\bsnm{Qiu}, \binits{W.}},
\bauthor{\bsnm{Yang}, \binits{J.}},
\bauthor{\bsnm{Wang}, \binits{B.}},
\bauthor{\bsnm{Yang}, \binits{M.}},
\bauthor{\bsnm{Tian}, \binits{G.}},
\bauthor{\bsnm{Wang}, \binits{P.}},
\bauthor{\bsnm{Yang}, \binits{J.}}:
\batitle{Evaluating the microsatellite instability of colorectal cancer based on multimodal deep learning integrating histopathological and molecular data}.
\bjtitle{Frontiers in Oncology}
\bvolume{12},
\bfpage{925079}
(\byear{2022})
\end{barticle}
\endbibitem

\bibitem[\protect\citeauthoryear{Alinia et~al.}{2024}]{Alinia2024}
\begin{botherref}
\oauthor{\bsnm{Alinia}, \binits{S.}},
\oauthor{\bsnm{Asghari-Jafarabadi}, \binits{M.}},
\oauthor{\bsnm{Mahmoudi}, \binits{L.}},
\oauthor{\bsnm{Roshanaei}, \binits{G.}},
\oauthor{\bsnm{Safari}, \binits{M.}}:
Predicting mortality and recurrence in colorectal cancer: Comparative assessment of predictive models.
Heliyon
\textbf{10}(6)
(2024)
\end{botherref}
\endbibitem

\bibitem[\protect\citeauthoryear{Fu et~al.}{2023}]{Fu2023}
\begin{barticle}
\bauthor{\bsnm{Fu}, \binits{X.}},
\bauthor{\bsnm{Patrick}, \binits{E.}},
\bauthor{\bsnm{Yang}, \binits{J.Y.}},
\bauthor{\bsnm{Feng}, \binits{D.D.}},
\bauthor{\bsnm{Kim}, \binits{J.}}:
\batitle{Deep multimodal graph-based network for survival prediction from highly multiplexed images and patient variables}.
\bjtitle{Computers in Biology and Medicine}
\bvolume{154},
\bfpage{106576}
(\byear{2023})
\end{barticle}
\endbibitem

\bibitem[\protect\citeauthoryear{Noman et~al.}{2025}]{noman2025leveraging}
\begin{barticle}
\bauthor{\bsnm{Noman}, \binits{S.M.}},
\bauthor{\bsnm{Fadel}, \binits{Y.M.}},
\bauthor{\bsnm{Henedak}, \binits{M.T.}},
\bauthor{\bsnm{Attia}, \binits{N.A.}},
\bauthor{\bsnm{Essam}, \binits{M.}},
\bauthor{\bsnm{Elmaasarawii}, \binits{S.}},
\bauthor{\bsnm{Fouad}, \binits{F.A.}},
\bauthor{\bsnm{Eltasawi}, \binits{E.G.}},
\bauthor{\bsnm{Al-Atabany}, \binits{W.}}:
\batitle{Leveraging survival analysis and machine learning for accurate prediction of breast cancer recurrence and metastasis}.
\bjtitle{Scientific Reports}
\bvolume{15}(\bissue{1}),
\bfpage{3728}
(\byear{2025})
\end{barticle}
\endbibitem

\bibitem[\protect\citeauthoryear{Chen et~al.}{2024}]{chen2024multimodal}
\begin{botherref}
\oauthor{\bsnm{Chen}, \binits{M.}},
\oauthor{\bsnm{Wang}, \binits{K.}},
\oauthor{\bsnm{Kapur}, \binits{P.}},
\oauthor{\bsnm{Brugarolas}, \binits{J.}},
\oauthor{\bsnm{Hannan}, \binits{R.}},
\oauthor{\bsnm{Wang}, \binits{J.}}:
A multimodal ensemble approach for clear cell renal cell carcinoma treatment outcome prediction.
arXiv preprint arXiv:2412.07136
(2024)
\end{botherref}
\endbibitem

\bibitem[\protect\citeauthoryear{Mahootiha et~al.}{2024}]{mahootiha2024multimodal}
\begin{barticle}
\bauthor{\bsnm{Mahootiha}, \binits{M.}},
\bauthor{\bsnm{Qadir}, \binits{H.A.}},
\bauthor{\bsnm{Bergsland}, \binits{J.}},
\bauthor{\bsnm{Balasingham}, \binits{I.}}:
\batitle{Multimodal deep learning for personalized renal cell carcinoma prognosis: Integrating ct imaging and clinical data}.
\bjtitle{Computer Methods and Programs in Biomedicine}
\bvolume{244},
\bfpage{107978}
(\byear{2024})
\end{barticle}
\endbibitem

\bibitem[\protect\citeauthoryear{Paverd et~al.}{2024}]{paverd2024radiology}
\begin{barticle}
\bauthor{\bsnm{Paverd}, \binits{H.}},
\bauthor{\bsnm{Zormpas-Petridis}, \binits{K.}},
\bauthor{\bsnm{Clayton}, \binits{H.}},
\bauthor{\bsnm{Burge}, \binits{S.}},
\bauthor{\bsnm{Crispin-Ortuzar}, \binits{M.}}:
\batitle{Radiology and multi-scale data integration for precision oncology}.
\bjtitle{NPJ Precision Oncology}
\bvolume{8}(\bissue{1}),
\bfpage{158}
(\byear{2024})
\end{barticle}
\endbibitem

\bibitem[\protect\citeauthoryear{Shi et~al.}{2023}]{Shi2023}
\begin{barticle}
\bauthor{\bsnm{Shi}, \binits{Y.}},
\bauthor{\bsnm{Olsson}, \binits{L.T.}},
\bauthor{\bsnm{Hoadley}, \binits{K.A.}},
\bauthor{\bsnm{Calhoun}, \binits{B.C.}},
\bauthor{\bsnm{Marron}, \binits{J.S.}},
\bauthor{\bsnm{Geradts}, \binits{J.}},
\bauthor{\bsnm{Niethammer}, \binits{M.}},
\bauthor{\bsnm{Troester}, \binits{M.A.}}:
\batitle{Predicting early breast cancer recurrence from histopathological images in the carolina breast cancer study}.
\bjtitle{NPJ Breast Cancer}
\bvolume{9}(\bissue{1}),
\bfpage{92}
(\byear{2023})
\end{barticle}
\endbibitem

\bibitem[\protect\citeauthoryear{Goyal et~al.}{2024}]{Goyal2024}
\begin{botherref}
\oauthor{\bsnm{Goyal}, \binits{M.}},
\oauthor{\bsnm{Marotti}, \binits{J.D.}},
\oauthor{\bsnm{Workman}, \binits{A.A.}},
\oauthor{\bsnm{Kuhn}, \binits{E.P.}},
\oauthor{\bsnm{Tooker}, \binits{G.M.}},
\oauthor{\bsnm{Ramin}, \binits{S.K.}},
\oauthor{\bsnm{Chamberlin}, \binits{M.D.}},
\oauthor{\bsnm{diFlorio-Alexander}, \binits{R.M.}},
\oauthor{\bsnm{Hassanpour}, \binits{S.}}:
Prediction of breast cancer recurrence risk using a multi-model approach integrating whole slide imaging and clinicopathologic features.
arXiv preprint arXiv:2401.15805
(2024)
\end{botherref}
\endbibitem

\bibitem[\protect\citeauthoryear{McInnes et~al.}{2018}]{mcinnes2018umap}
\begin{botherref}
\oauthor{\bsnm{McInnes}, \binits{L.}},
\oauthor{\bsnm{Healy}, \binits{J.}},
\oauthor{\bsnm{Melville}, \binits{J.}}:
Umap: Uniform manifold approximation and projection for dimension reduction.
arXiv preprint arXiv:1802.03426
(2018)
\end{botherref}
\endbibitem

\end{thebibliography}

\end{document}